\documentclass[
aps,%
12pt,%
final,%
notitlepage,%
oneside,%
onecolumn,%
nobibnotes,%
nofootinbib,
superscriptaddress,%
noshowpacs,%
centertags]%
{revtex4}
\usepackage[koi8-r]{inputenc}
\usepackage[T2A]{fontenc}
\usepackage[english,russian]{babel}
\usepackage{amsfonts}\usepackage{amsbsy}
\usepackage{feynmp}
\usepackage{graphicx}
\def\cal{\mathcal}
\newcommand{\di}{{\mathrm d}}
\newcommand{\Tr}{{\mathrm{Tr}}\,}
\newcommand{\ii}{{\mathrm i}}
\renewcommand{\Re}{{\mathrm{Re}}\,}
\renewcommand{\oint}{\int_{\cal C}}
\renewcommand{\Im}{{\mathrm{Im}}\,}
\def\Tc{{\cal T}_{\cal C}}
\def\scr#1{\mbox{\scriptsize #1}}
\def\vecs#1{\mbox{\scriptsize \boldmath $#1$}}
\def\vec#1{\mbox{\boldmath $#1$}}
\newcommand{\dpi}[1]{\frac{\di^4 #1}{(2\pi)^4}}                
\newcommand{\Pbr}[1]{\left\{#1\right\}}                    
\newlength{\charwidth}
\def\medhat#1{\settowidth{\charwidth}{$#1\,$}{\makebox[\charwidth]{$\,
 {\widehat{\makebox[2mm]{$#1\,$}}}$}}\vphantom{#1}}
\newcommand{\lap}%
{\;\raisebox{-0.5ex}{$\stackrel{\scriptstyle <}{\scriptstyle \sim}$}\;}
\newcommand{\gap}%
{\raisebox{-0.5ex}{$\stackrel{\scriptstyle >}{\scriptstyle \sim}$}}
\def\Gr{G}\def\Se{\Sigma}

\def\ja{\medhat{J}_a}\def\jad{\medhat{J}_a^\dagger}
\def\Pa{\medhat{\phi}_a}\def\Pad{\Pa^\dagger}
\def\Pb{\medhat{\phi}}
\def\Ps{\medhat{\psi}}\def\Psd{\Ps^\dagger}
\def\Ph{\medhat\phi}
\def\Pba{{\phi_a}}\def\jba{{J_a}}
\def\suma{\displaystyle {\sum_a}}
\def\mpa{\left(\mp\right)}
\def\Lg{{\cal L}}
\def\Lgh{\makebox[3.5mm]{${\widehat{\makebox[2mm]{$\Lg$}}}$}\vphantom{L}}
\def\Lint{\Lgh^{\mbox{\scriptsize int}}}
\def\A{A}
\def\Gm{\Gamma}
\def\F{F}                             
\def\Ft{\widetilde{F}}                 
\def\Ld{\Gamma^{\scr{out}}}
\def\Ldt{\Gamma^{\scr{in}}}
\def\Do{{\cal D}} 
\def\contourxy{\unitlength=0.8cm
\begin{picture}(17,1)\thicklines
\contour
\put(14.5,-.5){\makebox(0,0){$\infty$}}
\put(11,-.5){\makebox(0,0){$t_x^+$}}
\put(8,1.8){\makebox(0,0){$t_y^-$}}
\put(11,0){\circle*{.2}}
\put(8,1){\circle*{.2}}
\end{picture}}
\def\contour{\thicklines
\put(1,-.5){\makebox(0,0){$t_{0}$}}
\put(16.5,.5){\makebox(0,0){$t$}}
\put(0,.5){\vector(1,0){16}}
\put(1,1.){\line(1,0){13}}\put(14,.5){\oval(1,1)[br]}
\put(14,0){\vector(-1,0){13}}\put(14,.5){\oval(1,1)[tr]}}
\linethickness{1.0pt}

\begin{document}
\selectlanguage{english}  
\title{\vspace*{-15mm}
Self-consistent Approach to Off-Shell Transport\footnote{Dedicated to
  S.T. Belyaev on the occasion of his 80th birthday.}} 
\author{\firstname{Yu.B.~Ivanov}}
\email[]{Y.Ivanov@gsi.de}
\altaffiliation{RRC ``Kurchatov Institute'', Kurchatov sq.$\!$ 1, Moscow
123182, Russia}
\affiliation{Gesellschaft f\"ur Schwerionenforschung mbH, Planckstr.$\!$ 1,
64291 Darmstadt, Germany}
\author{\firstname{J. Knoll}}
\email[]{J.Knoll@gsi.de}
\affiliation{Gesellschaft f\"ur Schwerionenforschung mbH, Planckstr.$\!$ 1,
64291 Darmstadt, Germany}
\author{\firstname{D.N. Voskresensky}}
\email[]{D.Voskresensky@gsi.de}
\altaffiliation{Moscow Institute for Physics and Engineering, 
Kashirskoe sh. 31, Moscow 115409, Russia}
\affiliation{Gesellschaft f\"ur Schwerionenforschung mbH, Planckstr.$\!$ 1,
64291 Darmstadt, Germany}

\begin{abstract}
  The properties of two forms of the gradient expanded Kadanoff--Baym
  equations, i.e. the Kadanoff--Baym and Botermans--Malfliet forms,
  suitable to describe the transport dynamics of particles and
  resonances with broad spectral widths, are discussed in context of
  conservation laws, the definition of a kinetic entropy and the
  possibility of numerical realization.  Recent results 
  on exact conservations of charge and energy--momentum within
  Kadanoff--Baym form of quantum kinetics based on local coupling
  schemes are extended to two cases relevant in many applications.
  These concern the interaction via a finite range potential, and,
  relevant in nuclear and hadron physics, e.g. for the pion--nucleon
  interaction, the case of derivative coupling.
\end{abstract}
\maketitle
\begin{fmffile}{cons-fig}

\section{Introduction}\label{Intr}
Ever since L. Boltzmann had suggested his famous kinetic equation, the
field of non-equilibrium physics and stochastic processes has grown
tremendously, expanding into various directions. The interactions
among particles driven by mean fields were included,
quasiparticles were introduced in order to include much of the medium
effects, the kinematics was extended to the relativistic case,
ultimately theoretical foundations of transport equation were given
from an underlying quantum many-body or field theory.  In this line of
achievements also stands the work of G.I. Budker and S.T. Belyaev who
demonstrated the Lorentz invariance of the relativistic distribution
function and derived relativistic Fokker--Planck kinetic equation
\cite{BB56}. The work entered into many textbooks and found numerous
applications in atomic physics and electron-positron plasma.
Presently the relativistic transport concepts are a conventional tool
to analyze the dynamics of dense and highly excited matter produced in
relativistic heavy-ion collisions.

Along with semiphenomenological extensions a great progress was
achieved in microscopic foundation of the kinetic theory, which is
mainly associated with the names of Bogolyubov, Born, Green, Kirkwood,
Yvon, and Zubarev. The appropriate frame for description of
non-equilibrium processes within the real-time formalism of
quantum-field theory was developed by Schwinger, Kadanoff, Baym and
Keldysh \cite{Schw,Kad62,Keld64}. This formalism allows extensions of
the kinetic picture beyond conventional approximations (like the
quasiparticle one) and has found now numerous applications in many
domains of physics.

The interest in transport descriptions of heavy-ion collisions beyond
the quasiparticle approximation was initiated by Danielewicz
\cite{D2}, using the gradient expanded Kadanoff--Baym equations. These
attempts have recently has been revived
\cite{Knoll95,Bozek97,IKV99,IKHV00,KIV01,Cass99,Leupold,Mosel} in
order to properly describe the transport properties of broad resonances
(like the $\rho$ meson and $\Delta$ isobar). In the dense environment
also stable particle acquire a considerable width because of
collisional broadening. A proper dynamical treatment of their widths
in the dense nuclear medium within a transport theoretical concepts is
still a challenging problem.  Transport approaches for treating such
off-shell dynamics were proposed in refs.
\cite{Bozek97,IKV99,IKHV00,KIV01,Cass99,Leupold,Mosel}. They all were based
on the Kadanoff--Baym equations \cite{Kad62,D1} which describe the
non-equilibrium quantum evolution at the truncation level of the
Schwinger--Dyson equation.  Expanded up to the first space--time
gradients these provide equations for the one-body phase-space
distribution functions with collision term and Poisson-bracket terms
arising from the first order gradient terms.  Presently two
slightly different forms of the gradient-expanded Kadanoff--Baym
equations are used: the original Kadanoff--Baym (KB) form
\cite{Kad62}, as it follows right after the gradient expansion without
any further approximations, and the Botermans--Malfliet (BM) one
\cite{Bot90}, which is derived from the KB form by omitting certain
second-order space-time gradient corrections.

In this paper we would like to compare these two forms of ``quantum''
kinetic equations and discuss their advantages and disadvantages from
the point of view of their conserving properties, the possibility of
numerical realization, etc., sect. \ref{sect-Kin-EqT}. Technical
details on the conserving-properties are deferred to the appendices
(\ref{Der.-Coupl.}--\ref{Deriv}), since they illustrate some of the
general consideration of ref.  \cite{KIV01} together with some
extensions to cases particularly relevant in nuclear physics. Append.
\ref{Potent} treats nonrelativistic nucleon--nucleon interactions via
a potential of finite range. The derivative coupling is considered in
Append.  \ref{Deriv} at the example of the $P$-wave pion--nucleon
interaction.  In sect. \ref{Equilibration}, we supplement some
considerations about the construction of a kinetic entropy within
these two transport schemes. To make the paper self-contained, we
summarize the time-contour matrix notation in Append. \ref{Contour},
and introduce the $\Phi$-functional formalism for derivative coupling
in Append.  \ref{Der.-Coupl.}. A summary is given in sect.
\ref{Summary}.

\section{Off-Shell KB and BM Kinetics}
\label{sect-Kin-EqT}  

In this section we summarize the formulation of the off-shell
kinetic equations in the two different forms: in the 
KB form, i.e. as it follows right after the gradient
expansion of the exact Kadanoff--Baym equations, and in the 
BM form \cite{Bot90} which differs from the KB form only in the second order
of the gradient expansion. We assume the reader to be familiar with
the real-time formulation of non-equilibrium many-body theory and use
the contour matrix notation, detailed in Append. \ref{Contour}.

Starting point of all considerations is the set of Kadanoff-Baym
equations which express the space-time changes of the Wigner
transformed correlation function $\ii \Gr^{-+}(X,p)$ in terms of the
real-time contour convolution of the self-energy $\Se$ with the Green
function $\Gr$. We give the kinetic equation in compact notation (cf. Eq.
(\ref{H=FG}))
%
\begin{eqnarray}
\label{KB-eq.}
v_{\mu}\partial^{\mu}_X\ii\Gr^{-+}(X,p) = 
\left[ \Se\otimes\Gr - \Gr\otimes\Se \right]^{-+}_{X,p}
\quad\mbox{with}\quad
v^\mu=\frac{\partial}{\partial p_{\mu}}G_0^{-1}(p), 
\end{eqnarray}
%
where $\Gr^{-1}_{0}(p)$ is the Fourier transform of the inverse free Green
function 
\begin{eqnarray}
\label{G0}
\Gr^{-1}_{0}(p)=\left\{
\begin{array}{ll}
p^2-m^2\quad&\mbox{for relativistic bosons}\\
p_0-{\vec p}^2/(2m)\quad&\mbox{for non-rel. fermions or bosons.}
\end{array}\right.
\end{eqnarray}
For a complete definition, Eq. (\ref{KB-eq.}) has to be supplemented with
further equations, e.g., for the retarded Green function together with
the retarded relations (\ref{Fretarded}). In addition to these equations the
exact set of Kadanoff--Baym equations also includes the prototype of
the mass-shell equation, which we also discuss below. If a system
under consideration is only slightly spatially inhomogeneous and
slowly evolving in time, a good approximation is provided by an
expansion up to first order in space--time gradients.  Then the
main problem to arrive at a proper kinetic equation consists in 
accurately disentangling the rather complicated r.h.s. of Eq.
(\ref{KB-eq.}). This problem in the context of conserving
approximations will be addressed here.

\subsection{$\Phi$-derivable approximations}

In actual calculations one has to use certain approximations or
truncation schemes to the exact non-equilibrium theory, which make
conserving properties (such as charge and energy--momentum
conservations) and thermodynamic consistency of the transport theory
not evident.  It was shown \cite{Baym,IKV,IKV99} that there exist a
class of self-consistent approximations, called $\Phi$ derivable
approximations, which are conserving at the expectation value level
and at the same time thermodynamically consistent, i.e. they provide
true Noether currents and a conserved energy--momentum tensor. In
these schemes the self-consistent self-energies are generated from a
functional $\Phi[\Gr]$ through the following variational procedure
\cite{IKV99}
%
\begin{eqnarray}\label{var-Phi-component}
-\ii 
\Se_{ik}(X,p)=\mp\frac{\delta\ii\Phi[\Gr]}{\delta\ii \Gr^{ki}(X,p)}
\times \left\{
\begin{array}{ll}
2&\mbox{ for real fields}\\
1&\mbox{ for complex fields}
\end{array}\right. ,
\quad i,k\in\{-+\}. 
\end{eqnarray}
%
The functional $\Phi[\Gr]$ specifies the truncation scheme. It consists of
a set of properly chosen closed two-particle irreducible diagrams, where lines
denote the self-consistent propagator $\Gr$, while vertices are bare. The
functional variation with respect to $\Gr$ diagrammatically implies an opening
of a propagator line of $\Phi$. 

Particular examples of $\Phi$ derivable approximations can be found in 
Appendices \ref{Potent} and \ref{Deriv}, which consider applications
of the general formalism to cases important in nuclear physics. 
The treatment of the pion--nucleon derivative coupling in
Append. \ref{Deriv} requires the corresponding extension of the $\Phi$
derivable formalism, which has not been done up to now. Therefore, in 
Append. \ref{Der.-Coupl.} we perform such an extension and derive the 
relevant modifications of the variational rules and the ensuing additional
terms in the current and energy--momentum tensor expressions. 

The conserving properties of these approximations are {\em exact} at the
level of Kadanoff--Baym equations (\ref{KB-eq.}), while after the expansion up
to the first space--time gradients they are generally expected to be only {\em
  approximately} fulfilled.

\subsection{Physical notation}
It is helpful to eliminate the imaginary factors inherent in the standard
Green function formulation and introduce quantities which are real and, in the
quasi-homogeneous limit, positive, with clear physical meaning, thereby. Thus
instead of Green functions $\Gr^{ij}(X,p)$ and self-energies $\Se^{ij}(X,p)$
with $i,j\in\{-+\}$ (see Append. \ref{Contour}) in the Wigner representation
we  use the kinetic notation of ref. \cite{IKV99}.
We define\footnote{Here and below the upper sign corresponds to fermions,
while the lower sign, to bosons.}
%
\begin{eqnarray}
\label{F}
F (X,p) &=& \A (X,p)  f (X,p)
 =  (\mp )\ii \Gr^{-+} (X,p) , \nonumber\\
\Ft (X,p) &=& \A (X,p) [1 \mp  f (X,p)] = \ii \Gr^{+-} (X,p) , 
\end{eqnarray}
%
for the generalized Wigner functions $\F$ and $\Ft$ and the corresponding {\em
  four}-phase-space distribution functions $ f(X,p)$ and Fermi/Bose factors
$[1 \mp  f (X,p)]$. Here
%
\begin{eqnarray}
\label{A}
 A (X,p) \equiv -2\Im \Gr^R (X,p) = \Ft \pm F 
\end{eqnarray}
is the spectral function, and $\Gr^R$ is the retarded propagator. 
The spectral function satisfies the sum rule
%
\begin{eqnarray}
\label{A-sumf-nr} 
\int_{0}^{\infty} \frac{\di p_0}{2\pi} 
A(X,p) &=& 1 \quad \mbox{for nonrelativ. particles},
\\
\label{A-sumf-rb} 
\int_{-\infty}^{\infty} \frac{\di p_0}{2\pi} p_0
A(X,p) &=& 1 \quad \mbox{for relativ. bosons},
\end{eqnarray}
%
which follows from the canonical equal-time (anti) commutation relations
for (fermionic) bosonic field operators. 
Likewise the gain and loss rates of the collision integral are defined
as
%
\begin{eqnarray}
\label{gain}
\Ldt (X,p) =   \mp \ii \Se^{-+} (X,p),\quad 
\Ld (X,p)  =  \ii \Se^{+-} (X,p)  
\end{eqnarray}
%
with the damping width
%
\begin{eqnarray}
\label{G-def}
\Gm (X,p)&\equiv& -2\Im \Se^R (X,p) = \Ld (X,p)\pm\Ldt (X,p), 
\end{eqnarray}
%
where $\Se^R$ is the retarded self-energy. 

In terms of above kinetic notation, the gradient-expanded Kadanoff--Baym
equations are reduced to equations for four real quantities: two equations for
the real and imaginary parts of the retarded Green function, while there are
two equations for the phase-space occupation $F$: the KB kinetic equation and
the prototype ``mass-shell equation''. The latter doubling of equations
reflects the well known redundancy of the Kadanoff--Baym equations. Before the
gradient expansion both equations are completely identical. However, after the
gradient expansion their interrelation is no longer obvious and deserves
special care (see below).

The equations for the retarded propagator in first-order gradient
approximation can be immediately solved with the result
\cite{Kad62,Bot90}
%
\begin{eqnarray}
\label{Asol}
\Gr^R=\frac{1}{M(X,p)+\ii\Gm(X,p)/2}\Rightarrow
\left\{\begin{array}{rcl}
A (X,p) &=&\displaystyle
\frac{\Gm (X,p)}{M^2 (X,p) + \Gm^2 (X,p) /4}\\[4mm]
\Re\Gr^R (X,p) &=& \displaystyle 
\frac{M (X,p)}{M^2 (X,p) + \Gm^2 (X,p) /4}
\end{array}\right.
\end{eqnarray}
%
with the  ``mass'' function 
%
\begin{eqnarray}\label{M}
M(X,p)=\Gr^{-1}_{0}(p) -\Re\Se^R (X,p). 
\end{eqnarray}
Although the solution (\ref{Asol}) is simply algebraic, it is valid up to
first-order gradients.

%
%

\subsection{Kadanoff--Baym form}\label{KB}

In terms of above notation, the KB kinetic equation for $F$ in the
first-order gradient approximation takes the form
%
\begin{eqnarray}
\label{keqk1}
\Do 
F (X,p) - 
\Pbr{\Ldt , \Re\Gr^R}
&=& C (X,p) . 
\end{eqnarray}
%
We denote this as the {\em quantum transport equation in
the KB-choice}\footnote{If the system consists of several different
particle species, there is a set of coupled kinetic equations
corresponding to each species (e.g., see Append. \ref{Deriv}).}.  
Here the differential drift operator is defined as
%
\begin{eqnarray}\label{Drift-O}
\Do = 
\left(
v_{\mu} - 
\frac{\partial \Re\Se^R}{\partial p^{\mu}} 
\right) 
\partial^{\mu}_X + 
\frac{\partial \Re\Se^R}{\partial X^{\mu}}  
\frac{\partial }{\partial p_{\mu}}, 
\end{eqnarray}
%
and $\Pbr{...,...}$ denotes the four-dimensional Poisson bracket
%
\begin{eqnarray}
\label{[]} 
\Pbr{f(X,p) , \varphi(X,p)} = 
\frac{\partial f}{\partial p^{\mu}}  
\frac{\partial \varphi}{\partial X_{\mu}} 
- 
\frac{\partial f}{\partial X^{\mu}}  
\frac{\partial \varphi}{\partial p_{\mu}}, 
\end{eqnarray}
%
in covariant notation. Please note that now after gradient
approximation all quantities on the l.h.s. are to be taken in local
approximation, i.e. void of any further gradient terms. Thus, the
occurring self-energies are obtained evaluating the diagrams as in
momentum representation with the coordinates $X$ of all Greens
functions kept identical. The r.h.s. specifies the collision
term\footnote{See an example in Eq.  (\ref{HBF-C}).}
%
\begin{eqnarray}
\label{Coll(kin)}
C (X,p) =
\Ldt (X,p) \Ft (X,p) 
- \Ld (X,p) \F (X,p).
\end{eqnarray}
%
If the diagrams for the self-energy contain internal vertices, which
give rise to memory or non-local effects, the gain and loss rates
contain additional gradient terms which have to be constructed, e.g.,
according to the rules given in \cite{KIV01}. The resulting local part
of the collision term is charge (e.g., the baryonic number) and
energy--momentum conserving by itself
%
\begin{eqnarray}
\label{C-conser}
\mbox{Tr} \int \dpi{p} 
\left(
\begin{array}{lll}
e\\p^\mu
\end{array}
\right)
C^{\scr{loc}}=0. 
\end{eqnarray}
%
Here and below,  $e$ denotes the elementary charge, while
  $\Tr$ implies the sum over all possible internal degrees of freedom,
  like spin, and over possible particle species. We do not explicitly
  introduce the particle-species label to avoid overcomplication of
  equations.
In terms of a local functional
$\Phi^{\scr{loc}}$ the explicit form of the local collision term is
%
\begin{eqnarray}
\label{Coll-var} 
C^{\scr{loc}} (X,p) = 
 \frac{\delta\ii\Phi^{\scr{loc}}}{\delta\Ft(X,p)}\Ft(X,p)
-\frac{\delta\ii\Phi^{\scr{loc}}}{\delta\F(X,p)}\F(X,p) , 
\end{eqnarray}
%
cf. Eq. (\ref{var-Phi-component}).  In this paper we limit the
considerations to cases void of memory effects in this collision term.
The latter effects were studied in \cite{IKV99}.

Relation (\ref{C-conser}) permits us to derive the current 
%
\begin{eqnarray}
\label{KB-eff-currentk} 
j^{\mu}_{\scr{KB-eff}} (X) 
= e \mbox{Tr} \int \dpi{p}
\left[ 
v^{\mu} F (X,p) 
+
\Re\Se^R  \frac{\partial F}{\partial p_{\mu}}
- 
\Re\Gr^R \frac{\partial\Ldt}{\partial p_{\mu}}  
\right]. 
\end{eqnarray}
%
of a charge $e$ (e.g., the baryonic one) 
from the KB kinetic equation (\ref{keqk1}) which is conserved
%
\begin{eqnarray}
\label{KB-eff-cons} 
\partial_{\mu}j^{\mu}_{\scr{KB-eff}} (X) = 0.  
\end{eqnarray}
%
Note that this current formally differs from the true
Noether current 
%
\begin{eqnarray}
\label{c-new-currentk} 
j^{\mu} (X) 
= e \mbox{Tr} \int \dpi{p} v^{\mu} F (X,p)+j^{\mu}_{\scr{(der)}}(X), 
\end{eqnarray}
%
which follows right from the operator expression for this quantity,
cf. ref.  \cite{IKV99} and Appendix \ref{Der.-Coupl.}. The additional
term $j^{\mu}_{\scr{(der)}}$ appears only in the case of derivative
coupling, see Eq. (\ref{var-vec}). In view of the gradient
approximation employed, one could generally expect both currents to
differ beyond the validity range of the gradient approximation.
However, as demonstrated in detail in ref.  \cite{KIV01}, these two
currents are exactly equal for $\Phi$-derivable approximations, if a
consistent gradient expansion is performed also in the gain and loss
rates (\ref{gain}) of the collision term (\ref{Coll(kin)}). In this
case the exact conservation of the Noether current results from the
corresponding invariance of the $\Phi$ functional -- (Eq. (6.9) in
\cite{IKV}) -- which survives the gradient expansion
%
\begin{equation}
\label{invarJk}
e \mbox{Tr} \int \dpi{p} 
\left[
\Pbr{\Re\Se^R,F} 
- 
\Pbr{\Re\Gr^R,\Ldt}  + C
\right]
= \partial_\mu j^{\mu}_{\scr{(der)}}. 
\end{equation}
%
The latter relation is written down for the general case of memory or
non-local effects included in $C$. If such effects in $C$ are absent,
the collision term drops out of Eq. (\ref{invarJk}) according to Eq.
(\ref{C-conser}).

Within $\Phi$-derivable approximations also the conservation of
energy--momentum can be established, for local (point-like) couplings
providing a local energy-momentum tensor.
The $p^{\nu}$ weighted four-momentum integral of the KB kinetic
equation leads to the following consistency relation
\def\Fnb{\mbox{$\;\left(\frac{1}{2}\right)_{\scr{n.b.}}$}}
\cite{IKV99}\footnote{
Here in compliance with  Eq. (\ref{var-Phi-component}) we define the factor
$\Fnb=\left\{
\begin{array}{cl}
1/2&\mbox{for neutral bosons (real fields)}\\
1&\mbox{else}
\end{array}\right.$.}
%
\begin{eqnarray}
\label{epsilon-invk}
&&
\partial^{\nu}
\left(
{\cal E}^{\scr{pot}} - {\cal E}^{\scr{int}} 
\right)-\partial_\mu{\cal E}_{\scr{(der.)}}^{\mu\nu}
\cr &&
= \mbox{Tr}
\Fnb
\int 
\frac{p^\nu \di^4 p}{(2\pi )^4}
\left[
\Pbr{\Re\Se^R,F} 
- 
\Pbr{\Re\Gr^R,\Ldt}  + C
\right], 
\end{eqnarray}
%
which is again {\em exact} after the gradient expansion as shown in
ref. \cite{KIV01} (see also Appendices \ref{Potent} and \ref{Deriv}).
It implies that the Noether energy--momentum tensor
%
\begin{eqnarray}
\label{E-M-new-tensork}
\Theta^{\mu\nu}(X)
&=&
\mbox{Tr} 
\Fnb
\int \dpi{p} 
v^{\mu} p^{\nu} F (X,p)
+ g^{\mu\nu}\left(
{\cal E}^{\scr{int}}-{\cal E}^{\scr{pot}}
\right)+{\cal E}_{\scr{(der.)}}^{\mu\nu}
\end{eqnarray}
%
is {\em exactly} conserved by the kinetic equation (\ref{keqk1}) 
%
\begin{eqnarray}
\label{appr-cons} 
\partial_{\mu} \Theta^{\mu\nu}(X) = 0. 
\end{eqnarray}
%
Here potential energy density ${\cal E}^{\scr{pot}}(X)$, which a probe
particle with Wigner density $F(X,p)$ would experience due to the
interaction with all other particles in the system, is
%
\begin{eqnarray}
\label{eps-potk-K-B}
{\cal E}^{\scr{pot}}(X)
= 
\mbox{Tr}
\Fnb
\int\dpi{p} \left[
\Re\Se^R F
+ \Re\Gr^R \Ldt
\right].  
\end{eqnarray}
%
The interaction energy density ${\cal E}^{\scr{int}}(X)$ specifies that
part of the total energy density which is due to interactions. In simple
cases it relates to ${\cal E}^{\scr{pot}}$ by a simple counting factor,
namely, if all the interaction vertices of a theory have the same
number $n_{l}$ of lines attached to them
%
\begin{eqnarray}\label{int-spec}
{\cal E}^{\scr{int}}(X)=\frac{2}{n_{l}}{\cal E}^{\scr{pot}}(X).
\end{eqnarray}
%
In particular, for two-body interactions one has $n_{l}=4$
and thus ${\cal E}^{\scr{int}}=\frac{1}{2}{\cal E}^{\scr{pot}}$, while
for the fermion--boson interaction $n_{l}=3$, which results in ${\cal
  E}^{\scr{int}}=\frac{2}{3}{\cal E}^{\scr{pot}}$. In Appendices
\ref{Potent} and \ref{Deriv} we discuss  cases of this type. The
additional term ${\cal E}_{\scr{(der.)}}^{\mu\nu}$ appears in Eq.
(\ref{E-M-new-tensork}) only in the case of derivative coupling, cf.
Eq. (\ref{var-tens}).

The considerations given above summarize the results of ref.
\cite{KIV01} which are quite general. However they are restricted to
local (point-like) interactions and void of derivative couplings. This
excludes two important cases relevant to many areas in physics,
nuclear physics in particular. These are the cases of interaction
mediated by finite range non-relativistic potentials and of derivative
couplings like the $P$-wave pion--nucleon interaction.  Since the
considerations are rather technical they are exemplified in Appendices
\ref{Potent} and \ref{Deriv}. There the results of \cite{KIV01} are
generalized proving that also in these cases conserved currents and
expressions for a conserved total energy and total momentum can be
constructed. These two appendices, also provide further illustrations
of the discussion given in the present section.

The conserving feature are especially important for devising numerical
simulation codes based on this kinetic equation.  Indeed, if a
test-particle method is used, one should be sure that the number of
test particles is exactly conserved rather than approximately.  For a
direct application of this method, however, there is a particular
problem with the KB kinetic equation. In the test-particle method the
distribution functions are represented by an ensemble of test
particles as follows
%
\begin{eqnarray}
\label{test-part}
F (X,p) \sim \sum_i \delta^{(3)} \left({\bf X} - {\bf X}_i (T)\right)
\delta^{(4)} \left(p-p_i (T)\right),
\end{eqnarray}
%
where the $i$-sum runs over test particles. Then the $\Do F$ term in
Eq. (\ref{keqk1}) just corresponds to the classical motion of these
test particles subjected to forces inferred from $\Re \Se^{R}$, while
the collision term $C$ gives stochastic change of test-particles'
momenta, when their trajectories ``cross''. The additional term, i.e.
the Poisson-bracket term $\Pbr{\Ldt,\Re\Gr^R}$, spoils this simplistic
picture, since derivatives acting on the distribution function $F$
appear here only indirectly. Namely they are encoded through
derivatives of $\Ldt$. This term is responsible for backflow effects
which restore the Noether current to be the conserved one. However,
such backflow phenomena are difficult to absorb into test particles,
since they describe the response of the medium to the motion of the
charges.  In order to conserve the number of test particles between
subsequent collisions, one would have to unite the additional term
$\Pbr{\Ldt,\Re\Gr^R}$ with the drift term $\Do F$ even in the simplest
case, when the collision term is charge conserving by itself (see
Eq.(\ref{C-conser})) and derivative currents vanish,
$j^{\mu}_{\scr{(der)}}=0$.  However, the interpretation of the
additional term $\Pbr{\Ldt,\Re\Gr^R}$ causes problems within this
picture, since it is not just proportional to the same
$\delta$-functions as in Eq.  (\ref{test-part}) and thus can not be
included in the collisionless propagation of test particles. This
problem, of course, does not prevent a direct solution of the kinetic
equation. E.g., one can apply well developed lattice methods, which
are, however, much more complicated and time-consuming as compared to
the test-particles approach.

Within the same approximation level the set of Dyson equations for Green
functions $\Gr^{ij}(X,p)$ provides us with an alternative equation for $F$ 
%
\begin{eqnarray}
\label{mseq(k)1}
MF - \Re\Gr^R\Ldt
=\frac{1}{4}\left(\Pbr{\Gm,F} - \Pbr{\Ldt,\A}\right),
\end{eqnarray}
%
which is called the mass-shell equation, since in the quasiparticle
limit it provides the mass-condition $M=0$.  This equation coincides
with the kinetic one (\ref{keqk1}) only within the first-order
gradient approximation \cite{IKV99,Cass99,Leupold,Bot90}, while both
equations are {\em exactly} identical before the gradient expansion.
In view of this still remaining difference the practical recipe is to
forget about the mass-shell equation (\ref{mseq(k)1}), since the
retarded equation (\ref{Asol}) determines the spectral distribution, 
and to treat Eq.  (\ref{keqk1}) as a proper quantum kinetic equation.
Still this is an ambiguous recipe, which historically was one of the
motivation to proceed to the Botermans--Malfliet form of the quantum
kinetic equation.

\subsection{Botermans--Malfliet form}\label{BM}

As can be seen from the mass-shell equation (\ref{mseq(k)1}) and
Eq. (\ref{keqk1}) \cite{IKV99,Cass99,Leupold,Bot90}, the gain rate $\Ldt$
departs from $F\Gamma/A$ only by corrections of first order in the gradients
%
\begin{eqnarray}
\label{BM-subst}
\Ldt = \Gm F/\A + O(\partial_X)
\end{eqnarray}
%
(in equilibrium both equate each another). This fact permits to
substitute the r.h.s. estimate for $\Ldt$ in any of the gradient
terms, i.e. in the Poisson-bracketed terms of Eqs. (\ref{keqk1}) and
(\ref{mseq(k)1}) and neglect the correction $O(\partial_X)$ as it
leads to terms of already second-order in the gradients. Upon this
substitution, first proposed by Botermans and Malfliet \cite{Bot90},
one arrives at the following form of the kinetic and mass-shell
equations
%
\begin{eqnarray}
\label{keqk}
\Do 
F (X,p) - 
\Pbr{\Gm\frac{F}{\A},\Re\Gr^R} &=& C (X,p), 
\end{eqnarray}
%
%
\begin{eqnarray}
\label{mseq(k)}
MF - \Re\Gr^R\Ldt
=\frac{1}{4}\left(\Pbr{\Gm,F} - \Pbr{\frac{\Gm F}{\A},\A}\right),
\end{eqnarray}
%
which are already {\em exactly} identical, as they were before the gradient
expansion, and still equivalent to those in the KB form within the first-order
gradient approximation. The so obtained equation (\ref{keqk}) will be called 
{\em quantum kinetic equations  in BM-choice}. 
This equation {\em exactly} conserves the following effective current 
%
\begin{eqnarray}
\label{BM-eff-currentk} 
j^{\mu}_{\scr{BM-eff}} (X) 
= e \mbox{Tr} \int \dpi{p}
\left[ 
v^{\mu} F (X,p) 
+
\Re\Se^R  \frac{\partial F}{\partial p_{\mu}}
- 
\Re\Gr^R \frac{\partial(\Gm F/\A)}{\partial p_{\mu}}  
\right], 
\end{eqnarray}
%
which differs from the Noether current $j^{\mu}$ in terms of the order
of $O(\partial_X)$, provided a $\Phi$-derivable approximation is used
for self-energies. All KB-choice properties of Eq. (\ref{keqk}) within
a $\Phi$-derivable approximation also transcribe to BM-choice through
the substitution $\Ldt=\Gm F/\A$ in Eqs. (\ref{invarJk}),
(\ref{epsilon-invk}) and (\ref{eps-potk-K-B}). This substitution,
however, touches the accuracy of those relations. For instance, the
conservation laws of the Noether currents (\ref{c-new-currentk}) and
the energy--momentum tensor (\ref{E-M-new-tensork}) are then only
approximately fulfilled together with the corresponding consistency
relations (\ref{invarJk}) and (\ref{epsilon-invk}) which now look as
%
\begin{equation}
\label{invarJk-BM}
e \mbox{Tr} \int \dpi{p} 
\left[
\Pbr{\Re\Se^R,F} 
- 
\Pbr{\Re\Gr^R,\Gm F/\A}  + C
\right]
\simeq \partial_\mu j^{\mu}_{\scr{(der)}}, 
\end{equation}
%
%
\begin{eqnarray}
\label{epsilon-invk-BM}
&&
\partial^{\nu}
\left(
{\cal E}^{\scr{pot}} - {\cal E}^{\scr{int}} 
\right)-\partial_\mu{\cal E}_{\scr{(der.)}}^{\mu\nu}
\cr &&
\simeq \mbox{Tr}
\left(\frac{1}{2}\right)_{\scr{n.b.}}
\int 
\frac{p^\nu \di^4 p}{(2\pi )^4}
\left[
\Pbr{\Re\Se^R,F} 
- 
\Pbr{\Re\Gr^R,\frac{\Gm F}{\A}}  + C
\right], 
\end{eqnarray}
%
respectively, and hold only up to first-order gradients.  

The effective BM-current (\ref{BM-eff-currentk}) was used by
S. Leupold \cite{Leupold} as a basis for the construction of a
test-particle ansatz for the nonrelativistic case. In this case the
additional term $\Pbr{\Gm F/\A,\Re\Gr^R}$ in the BM kinetic equation
(\ref{keqk}) is expressed in terms of the same distribution function
as the drift term $\Do F$. Therefore, one can unify these terms
to construct equations of motions for test particles, which provide
exact conservation of $j^{\mu}_{\scr{BM-eff}}$. To automatically
fulfill this effective current conservation, the test-particle ansatz
is introduced for the combination
%
\begin{eqnarray}
\label{test-part-L}
\frac{1}{2} \Gm\A\left(1-\frac{\partial \Re\Se^R}{\partial p_0}
- \frac{M}{\Gm} \frac{\partial \Gm}{\partial p_0}\right) 
F (X,p)
\sim \sum_i \delta^{(3)} \left({\bf X} - {\bf X}_i (T)\right)
\delta^{(4)} \left(p-p_i (T)\right),
\end{eqnarray}
%
rather than for the distribution function itself. Note that the energy
$p^0_i(T)$ of the test particle is a free coordinate, not restricted
my a mass-shell condition. W. Cassing and S. Juchem \cite{Cass99}
extended this test-particle ansatz to the relativistic case.  The
equations of motion for the test particle, which follow from this
ansatz, in particular give the time evolution of the off-shellness of
a test particle \cite{Cass99,Leupold}
%
\begin{eqnarray}
\label{off-shellness}
\frac{\di M}{\di T}=\frac{M}{\Gm}\frac{\di \Gm}{\di T}, 
\end{eqnarray}
%
the origin of which can be traced back to the additional term
$\Pbr{\Gm F/\A,\Re\Gr^R}$ in the BM kinetic equation (\ref{keqk}).
Here $M$ is the mass of the test particle relative to its on-shell
value, see Eq. (\ref{M}), and this equation of motion implies that
once the width drops in time the particles are driven towards the
on-shell mass $M=0$. This clarifies the meaning of the letter term for the
off-shell BM transport: it provides the time evolution of the
off-shellness.

\section{Entropy}\label{Equilibration}

Another important feature of the kinetic description is the approach
to thermal equilibrium during evolution of a closed system. In terms
of transport theory the sufficient (while not necessary!) condition of
it is the existence of an H-theorem. Leaving aside all complications
associated with non-local effects in the collision term and possible lack
of positive definiteness of the transition rates, discussed in ref.
\cite{IKV99}, we confine our consideration to simple
approximations, cf. (\ref{V-phi}). As demonstrated in ref.
\cite{IKV99}, in the BM approximation to the quantum kinetic equations
(\ref{keqk}) the H-theorem can indeed be formulated
%
\begin{eqnarray}
\label{H_BM} 
\partial_\mu s^\mu_{\scr{BM}} (X) =\mbox{Tr}
\int \dpi{p} \ln \frac{\Ft}{\F} C^{\scr{loc}} (X,p) \geq 0,     
\end{eqnarray}
%
where the quantity 
%
\begin{eqnarray}
\label{S_BM} 
s^\mu_{\scr{BM}} =\mbox{Tr}
\int \dpi{p}
\left[
\left(
v^{\mu} - \frac{\partial \Re\Se^R}{\partial p_{\mu}}
\right)
\left(
\mp \Ft \ln \frac{\Ft}{A} - \F \ln \frac{\F}{A} 
\right)
\right.\;\,&&
\nonumber 
\\
- 
\left. 
\Re\Gr^R
\left(
\mp \ln\frac{\Ft}{A} \frac{\partial}{\partial p_{\mu}} 
\left(\Gm\frac{\Ft}{\A} \right)
- 
\ln\frac{\F}{A} \frac{\partial}{\partial p_{\mu}} 
\left(\Gm\frac{\F}{\A} \right)
\right)
\right]&& 
\end{eqnarray}
%
obtained from the l.h.s. of the BM kinetic equation (\ref{keqk}) is
interpreted as a entropy flow for the BM-choice. For the
$\Phi$-derivable approximation (\ref{V-phi}) the r.h.s. of relation
(\ref{H_BM}) takes the following form 
%
\begin{eqnarray}
\label{s(coll)} 
&&\mbox{Tr}\int \dpi{p} \ln \frac{\Ft}{\F} C^{\scr{loc}} (X,p) 
= \mbox{Tr}
\frac{1}{4} 
 \int \dpi{p_1}\dpi{p_2} \dpi{p_3}\dpi{p_4}
\nonumber 
\\
&&\times
R^{\scr{loc}}\; (2\pi)^4\delta^4\left(p_1 + p_2 - p_3 - p_4\right)
\left(\F_1\F_2 \Ft_3\Ft_4 
-
\Ft_1\Ft_2 \F_3\F_4\right)
\ln\frac{\F_1 \F_2 \Ft_3\Ft_4}
        {\Ft_1\Ft_2\F_3\F_4}, 
\end{eqnarray}
%
where $R^{\scr{loc}}$ is the transition rate determined by Eq.
(\ref{gain-V}).  This expression is indeed non-negative, since
$(x-y)\ln(x/y) \ge 0$ for any positive $x$ and $y$, and is of the
second order in deviation from equilibrium $(\F-\F_{\scr{eq}})$, as
both $(\F_1\F_2\Ft_3\Ft_4-\Ft_1\Ft_2\F_3\F_4)$ and
$\ln(\F_1\F_2\Ft_3\Ft_4/\Ft_1\Ft_2\F_3\F_4)$ approach zero at
equilibrium. From the kinetic equation it follows that the deviation from
equilibrium is of the first order in time gradients:
$(\F-\F_{\scr{eq}})\propto O(\partial_T \F)$. This implies that the
r.h.s. of relation (\ref{H_BM}) is of the second order in time
gradients, which is, strictly speaking, beyond our first-order
gradient approximation. However, from the point of view of practical
use this feature is highly welcome as it guarantees equilibration.  A
further advantaged of the kinetic entropy flux (\ref{S_BM}) is that in
equilibrium its entropy merges that thermodynamic expression deduced from the
thermodynamic potential in the $\Phi$ derivable scheme 
\cite{IKV99,Norton,Carneiro}.

In the case of the KB-choice (\ref{keqk1}) the situation is more
controversial. Performing all the same manipulations with the KB kinetic
equation (\ref{keqk1}) as those in ref. \cite{IKV99}, we arrive at the
following relation 
%
\begin{eqnarray}
\label{H_KB} 
\partial_\mu s^\mu_{\scr{KB}} (X) =
\Tr \int \dpi{p} \ln \frac{\Ft}{\F} C^{\scr{loc}} 
-\delta H_{\scr{KB}},   
\end{eqnarray}
%
where 
%
\begin{eqnarray}
\label{S_KB} 
s^\mu_{\scr{KB}} =\Tr
\int \dpi{p}
\left[
\left(
v^{\mu} - \frac{\partial \Re\Se^R}{\partial p_{\mu}}
\right)
\left(
\mp \Ft \ln \frac{\Ft}{A} - \F \ln \frac{\F}{A} 
\right)
\right.\;\,&&
\nonumber 
\\
- 
\left. 
\Re\Gr^R
\left(
\mp \ln\frac{\Ft}{A} \frac{\partial\Ld}{\partial p_{\mu}} 
- 
\ln\frac{\F}{A} \frac{\partial\Ldt}{\partial p_{\mu}} 
\right)
\right].&& 
\end{eqnarray}
%
\begin{eqnarray}\label{cor}
\delta H_{\scr{KB}} =-
\int \dpi{p} \Re \Gr^R 
\Pbr{\ln \frac{\Ft}{\F}, \frac{C^{\scr{loc}}}{A}}.
\end{eqnarray}
The KB entropy flow 
$s^\mu_{\scr{KB}}$  is identical to the BM one
$s^\mu_{\scr{BM}}$ up to zero-order gradients, while  
they differ in the first-order gradient corrections.  One can easily 
obtain $s^\mu_{\scr{BM}}$ from the KB entropy flow
by doing replacement (\ref{BM-subst}) in $\Ldt$ 
and a similar replacement in $\Ld$.

The additional term $\delta H_{\scr{KB}}$ on the r.h.s. of relation
(\ref{H_KB}) is of the second order in gradients, due to the Poisson
bracket and $C^{\scr{loc}}\propto O(\partial_T F)$.  Therefore, the
r.h.s. of (\ref{H_KB}) consists of two terms, which are of the same
order of magnitude and one of them ($\delta H_{\scr{KB}}$) is sign
indefinite. This prevents us from concluding the positive definiteness
of the r.h.s. of Eq. (\ref{H_KB}). Alternatively we were not able to
cast this term into a full divergence as to be included into the
definition of the KB entropy flow. This fact by itself does not
imply that the system does not approach equilibrium or even the
absence of an H-theorem for the KB kinetic equation but suggests that
equilibration should be tested in actual calculations. The local
H-theorem we are looking for is a very stringent condition, providing
monotonous approach to equilibrium. In fact, equilibration may well
be nonmonotonous in time.

Still, for the KB kinetic equation we are able to prove the H-theorem
in a limiting case, i.e. close to local thermal equilibrium or for 
quasi-stationary state, which slowly evolve in space and time. To be
definite, let us talk about the local thermal equilibrium.  In terms
of the distribution function
%
\begin{eqnarray}
\label{F+dF}
F (X,p)=F_{\scr{loc.eq.}} (X,p)+\delta F (X,p)
\end{eqnarray}
%
the above assumption implies that $|\delta F| \ll F_{\scr{loc.eq.}}$ and 
$|\partial_X F_{\scr{loc.eq.}}|  \lap  |\partial_X \delta F|$. 
Then we can write down 
\begin{eqnarray}
\label{dHKB}
\delta H_{\scr{KB}}=
\partial_{\mu}\delta s^{\mu}_{\scr{KB}} (X)+ 
\int \dpi{p} \frac{C^{\scr{loc}}}{A} 
\Pbr{\ln \frac{\Ft}{\F},\Re \Gr^R}.
\end{eqnarray}
where 
\begin{eqnarray}
\label{dsKB}
\delta s^{\mu}_{\scr{KB}} (X)= - \mbox{Tr}
\int \dpi{p} \frac{\Re\Gr^R}{A}
\frac{\partial \ln(\Ft/\F)}{\partial p_\mu}
C^{\scr{loc}}(X,p). 
\end{eqnarray}
Here, the remaining term
\begin{eqnarray}
\label{dHKB-rem}
\int \dpi{p} \frac{C^{\scr{loc}}}{A} 
\Pbr{\ln \frac{\Ft}{\F},\Re \Gr^R}
\propto 
O(\delta F\partial_X \delta F)+ 
O(\delta F\partial_X F_{\scr{loc.eq.}}), 
\end{eqnarray}
can be neglected, as it has additional gradient smallness
as compared to the first term on the r.h.s. of
Eq. (\ref{H_KB}). Here, we have taken into account that 
$C^{\scr{loc}} \propto \delta F$ and 
$\Pbr{\ln(\Ft/\F),\Re \Gr^R}\propto \partial_X (F_{\scr{loc.eq.}}+\delta F)$. 
Thus, from Eq. (\ref{H_KB}) we conclude that 
%
\begin{eqnarray}
\label{H_KB-teor} 
\partial_\mu \left(s^\mu_{\scr{KB}} +
\delta s^{\mu}_{\scr{KB}}\right) \geq 0 \quad
\mbox{near local equilibrium},  
\end{eqnarray}
%
which is the H-theorem for this case with the total entropy flow 
$s^{\mu}_{\scr{KB}}+\delta s^{\mu}_{\scr{KB}}$. 
Note that $\delta s^{\mu}_{\scr{KB}}$ is proportional to the collision
term and hence equals zero in equilibrium. 
The applicability range of this result is the same as that
for the memory entropy derived in \cite{IKV99} for the BM-choice.

\section{Summary and Perspectives}\label{Summary}

In conclusion, we would like to summarize the present status of
the two considered approaches to  off-shell transport. 

From a consistency point of view, the BM-choice looks more appealing, since it
preserves the {\em exact} identity between the kinetic and mass-shell
equation, a property inherent in the original KB equations  \cite{IKV99}.
For the KB-choice this identity between the kinetic and mass-shell equation is
only {\em approximately} preserved, namely within the validity range of the
first-order gradient approximation. However, this disadvantage is not of great
practical use, since, in any case, only one of these two equations, namely the
kinetic one, should be used in actual calculations.

For the construction of conservation laws related to global symmetries or
energy and momentum the {\em local} collision term entirely drops out of the
balance. Thus, the conservation laws solely depend on the properties of the
first order gradient terms in the kinetic equation. In this respect the
KB kinetic equation has a conceptual advantage as it leads to
{\em exact} \cite{KIV01} rather than approximate conservation laws, provided
the scheme is based on $\Phi$-derivable approximations.  Thereby the
expectation values of the original operator expressions of conserved quantities
(e.g., Noether currents) are exactly conserved.  The reason is that the
KB kinetic equation preserves certain contour symmetries among the various
gradient terms, while they are violated for the BM-choice. Of course, within
their range of applicability these two approaches are equivalent, because the
BM kinetic equation conserves the charge and energy--momentum within the
theoretical accuracy of the gradient approximation. Still, the fact that the
KB-choice posses exact conservation laws put this version to the level of a
generic equation, much like the Boltzmann or hydrodynamic equations, to be
used as phenomenological dynamical equations for practical applications.
Such conserving dynamical schemes may be useful even though the applicability
condition of the approximation might be violated at some stages of evolution.
For instance, such a situation happens at the initial stage of heavy-ion
collisions. As the conservations are exact, we can still use the gradient
approximation, relying on a minor role of this rather short initial stage in
the total evolution of a system. Moreover, exact conservation laws allow us to
keep control of numerical codes.

Although the KB kinetic equations posses exactly conserved Noether currents, a
practical numerical approach (e.g., by a test-particle method) for its solution
has not yet been established. The obstacle is the special Poisson-bracket term
in the KB kinetic Eq. (\ref{keqk1}) which lacks proper interpretation since the
phase-space occupation function $F(X,p)$ enters only indirectly through the
gain-rate gradient terms. What is known is that this term encodes the
backflow component which ensures the Noether currents to be conserved.
However such backflow features are difficult to be implemented into a 
test-particle scheme.  This problem, of course, does not exclude
solution of the KB 
kinetic equation, e.g within well adapted lattice methods, which are, however,
much more complicated and time-consuming as compared to the test-particle
approach.  For the BM kinetic equation, on the other hand, an efficient
test-particle method is already available \cite{Cass99,Leupold}, for
the price that 
it deals with an alternative current rather than the Noether current.

As a novel part we showed (cf. appendices) that the exact conservation
laws in the KB kinetic equations, 
originally derived for local interaction terms which lead to a local energy
momentum tensor, also do hold for derivative couplings and for interactions of
finite range, like a non-relativistic potential. For the latter case, of
course, only global conservation of energy and momentum can be achieved.
In order to deal with the derivative coupling, we extended the
$\Phi$-derivable approach to this case.

An important feature of kinetic descriptions is the approach to
thermal equilibrium during evolution of a closed system. A sufficient
(while not necessary!) condition is provided by an H-theorem. As was
demonstrated in ref.  \cite{IKV99}, at least, within simplest
$\Phi$-derivable approximations for the kinetic equation in BM-choice
an H-theorem indeed can be proven. The so derived kinetic entropy
merged the equilibrium expression which in the context of
$\Phi$-derivable approximations results from the thermodynamic
potential, cf. \cite{IKV99,Norton,Carneiro}.  For the KB kinetic
equation the result is by far weaker. Here we were able to prove the
H-theorem only within simplest $\Phi$-derivable approximations and for
a system very close to almost spatially homogeneous thermal local
equilibrium or stationary state.  These results, in general, do not
imply that the system does not approach equilibrium but suggests that
equilibration should be tested in actual calculations. Furthermore,
the local H-theorem with a local entropy current which we considered
for the BM-case may be by far too restrictive, providing monotonous
approaching to equilibrium. In fact, for kinetic equations with memory
or non-local effects equilibration may well be nonmonotonous in time.

Though the discussion in this paper is confined to problems of
$\Phi$-derivable off-shell transport based on the first-order gradient
expansion, essential progress has recently been achieved also in
solving KB equations directly without any gradient expansion for
selected examples. These concern non-equilibrium processes in
scalar and spinor--scalar models on 1 and 3 space dimensions, see e.g.
\cite{Berg01,Berg02a,Berg02b} and references 
therein.  It was found \cite{Berg01,Berg02a} that after a comparably
short but violent non-equilibrium evolution the time dependence of the
Wigner transformed spectral function becomes rather weak even for
moderate coupling constants. During this slow evolution the system is
still far away from equilibrium. This fact provides a necessary
condition for a successful gradient expansion and hence indicates a
wide range of applicability of the approaches discussed in this paper.
Even though the rapid far-from-equilibrium dynamics is formally beyond
the scope of applicability of the gradient-expanded quantum kinetics,
nevertheless, the KB choice includes all the ingredients required for
such a treatment, i.e. the proper mean-field dynamics, together with
the off-shell transport of particles, thereby satisfying exact rather
than approximate conservation laws even far away from equilibrium.

Further progress in understanding the properties of $\Phi$-derivable
approximations to finite-temperature quantum field theory was reported
concerning the question of renormalizabilty.  The new results are
equally applicable to quantum kinetic equations, both in KB or BM
choices. In ref.  \cite{KH02a} it was shown that truncated
non-perturbative self-consistent Dyson resummation schemes can be
renormalized with local counter terms defined at the vacuum level.
The requirements are that the underlying theory is renormalizable and
that the self-consistent scheme follows Baym's $\Phi$-derivable
concept.  This result proves that there is no arbitrariness in
studying the in-medium modifications of model parameters like the mass
and the coupling constants within this class of approximation schemes.
It is sufficient to adjust them in the vacuum, for instance, by
fitting them to scattering data, in order to predict their changes in
medium without ambiguity. This result also guarantees the standard
$\Phi$-derivable properties like thermodynamic consistency and exact
conservation laws also for the renormalized approximation schemes to
hold. In ref.  \cite{KH02b} the theoretical concepts for the
renormalization, devised in \cite{KH02a}, were applied to the
$\phi^4$-model, demonstrating the practicability of the method.

In general, the symmetries of the classical action which lead to
Ward-Takahashi identities for the proper vertex-functions are violated
for $\Phi$-derivable approximations for functions beyond the one-point
level, i.e., on the correlator level.  This causes problems concerning
the Nambu--Goldstone modes \cite{baymgrin} in the broken symmetry case
or concerning local symmetries (gauge symmetries) \cite{KH02c} on a
level where the gauge fields are treated beyond the classical field
approximation, i.e. on the propagator level. In ref. \cite{KH02c} it was
shown that on top of any solution of a $\Phi$-derivable approximation
which is constructed from a symmetric Lagrangian there exists a
non-perturbative effective action which generates proper vertex
functions in the same sense as the 1PI effective action. These
external vertex functions fulfill the Ward--Takahashi identities of the
underlying symmetry. However, in general they coincide with the
self-consistent ones only up to one-point order. Thus, usually the so
generated external self-energy and higher vertex functions are
different from the $\Phi$-derivable expressions. Therefore, the
pleasant property of the $\Phi$-derivable approximations, namely the 
conserving one, proves to be lost.  The derivation of approximation schemes
that fulfill all symmetry properties of the underlying classical
action and at the same time are fully self-consistent and conserving
still remains as an open task.

As has been already mentioned, the gauge invariance may be lost in
$\Phi$-derivable approximations, too. In particular, this problem
prevents applications of $\Phi$-derivable approximations (including
kinetic ones) to description quark--gluon plasma, based on QCD.  This
occurs because, in general, solution for dressed propagators and
vertices do not satisfy Ward--Takahashi identities. This pathology
shows up as an explicit dependence of results on the choice of the
gauge condition. In ref. \cite{Arriz} it was demonstrated, in fact,
that $\Phi$-derivable approximations have a controlled gauge
dependence, i.e. the gauge dependent terms appear at orders higher
than the truncation order. Furthermore, using the stationary point
obtained for the approximation to evaluate the complete 2PI effective
action boosts the order, at which the gauge dependent terms appear, to
twice the order of truncation. This is still not a solution of the
gauge problem in the rigorous sense but certain progress to its
better control and understanding.

\section*{Acknowledgments}
We are grateful to G. Baym, J. Berges, P. Danielewicz, H. Feldmeier,
B. Friman, H. van Hees, C. Greiner, E.E. Kolomeitsev and S. Leupold
for fruitful discussions on various aspects of this research. 
This work was supported in part by the Deutsche Forschungsgemeinschaft
(DFG project 436 RUS 113/558/0-2), the Russian Foundation for Basic
Research (RFBR grants 03-02-04008 and 00-15-96590) and the German BMBF
(contract RUS-01/690).

\appendix

\section{Matrix Notation} \label{Contour}

In calculations that apply the Wigner transformations, it is necessary to
decompose the full contour into its two branches---the {\em time-ordered} and
{\em anti-time-ordered} branches. One then has to distinguish between the
physical space-time coordinates $x,\dots$ and the corresponding contour
coordinates $x^{\cal C}$ which for a given $x$ take two values
$x^-=(x^-_{\mu})$ and $x^+=(x^+_{\mu})$ ($\mu\in\{0,1,2,3\}$) on the two
branches of the contour (see figure 1). \\ 
\begin{center}\vspace*{0.1cm}
\contourxy\\[0.5cm]
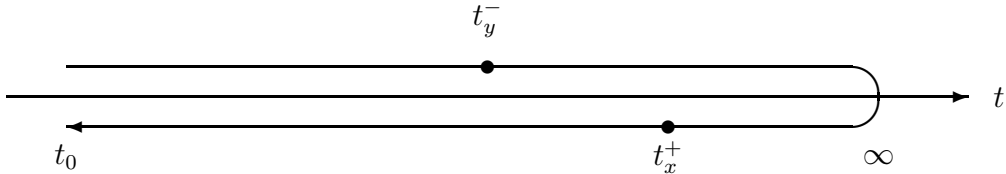
{\small Figure 1: Closed real-time contour with two external
  points $x,y$ on the contour.} 
\end{center}

Closed real-time contour integrations
can then be decomposed as
%
\begin{eqnarray}
\label{C-int}
\oint\di x \dots =\int_{t_0}^{\infty}\di x\dots
+\int^{t_0}_{\infty}\di x\dots
=\int_{t_0}^{\infty}\di x\dots -\int_{t_0}^{\infty}\di x\dots, 
\end{eqnarray}
%
where only the time limits are explicitly given.  The extra minus sign of the
anti-time-ordered branch can conveniently be formulated by a $\{-+\}$
``metric'' with the metric tensor in $\{-+\}$ indices
%
\begin{eqnarray}
\label{sig}
\left(\sigma^{ij}\right)&=&
\left(\sigma_{ij}\vphantom{\sigma^{ij}}\right)=
{\footnotesize\Big(\begin{array}{cc}1&0\\[-3mm] 
0& -1\end{array}\Big)}
\end{eqnarray}
%
which provides a proper matrix algebra for multi-point functions on the
contour with ``co''- and ``contra''-contour values.  Thus, for any two-point
function $F$, the contour values are defined as
%
\begin{eqnarray}\label{Fij}
F^{ij}(x,y)&:=&F(x^i,y^j), \quad i,j\in\{{\scriptstyle -\,,\, +}\},\quad\mbox{with}\cr
F_i^{~j}(x,y)&:=&\sigma_{ik}F^{kj}(x,y),\quad
F^i_{~j}(x,y):=F^{ik}(x,y)\sigma_{kj}\cr
F_{ij}(x,y)&:=&\sigma_{ik}\sigma_{jl}F^{kl}(x,y),
\quad\sigma_i^k=\delta_{ik}
\end{eqnarray}
%
on the different branches of the contour. Here summation over repeated indices
is implied. Then contour folding of contour two-point functions, e.g., in Dyson
equations, simply becomes
%
\begin{eqnarray}\label{H=FG}
H(x^i,y^k)&=&H^{ik}(x,y)= \left[ F\otimes G \right]^{ik}
\cr 
&\equiv& \oint\di z F(x^i,z)G(z,y^k)
=\int\di z F^i_{~j}(x,z)G^{jk}(z,y)
\end{eqnarray}
%
in the matrix notation.

For any multi-point function the external point $x_{max}$, which has the
largest physical time, can be placed on either branch of the contour without
changing the value, since the contour-time evolution from $x_{max}^-$ to
$x_{max}^+$ provides unity. Therefore, one-point functions have the same value
on both sides on the contour.

Due to the change of operator ordering, genuine multi-point functions are, in
general, discontinuous, when ever two contour coordinates become identical. In
particular, two-point functions like $\ii F(x,y)=\left<\Tc {\widehat
    A(x)}\medhat{B}(y)\right>$ become\footnote{Frequently used
  alternative notation is $F^{<}=F^{-+}$ and $F^{>}=F^{+-}$.}
%
\begin{eqnarray}\label{Fxy}
\hspace*{-0.5cm}\ii F(x,y) &=&
\left(\begin{array}{ccc} 
\ii F^{--}(x,y)&&\ii F^{-+}(x,y)\\[3mm]
\ii F^{+-}(x,y)&&\ii F^{++}(x,y)
\end{array}\right)=
\left(\begin{array}{ccc} 
\left<{\cal T}\medhat{A}(x)\medhat{B}(y)\right>&\hspace*{5mm}&
\mp \left<\medhat{B}(y)\medhat{A}(x)\right>\\[5mm]
\left<\medhat{A}(x)\medhat{B}(y)\right>
&&\left<{\cal T}^{-1}\medhat{A}(x)\medhat{B}(y)\right>
\end{array}\right), 
\end{eqnarray}
%
where ${\cal T}$ and ${\cal T}^{-1}$ are the usual time and anti-time ordering
operators.  Since there are altogether only two possible orderings of the two
operators, in fact given by the Wightman functions $F^{-+}$ and $F^{+-}$,
which are both continuous, not all four components of $F$ are independent. Eq.
(\ref{Fxy}) implies the following relations between nonequilibrium and usual
retarded and advanced functions
%
\begin{eqnarray}\label{Fretarded}
F^R(x,y)&=&F^{--}(x,y)-F^{-+}(x,y)=F^{+-}(x,y)-F^{++}(x,y)\nonumber\\
&:=&\Theta(x_0-y_0)\left(F^{+-}(x,y)-F^{-+}(x,y)\right),\nonumber\\
F^A (x,y)&=&F^{--}(x,y)-F^{+-}(x,y)=F^{-+}(x,y)-F^{++}(x,y)\nonumber\\
&:=&-\Theta(y_0-x_0)\left(F^{+-}(x,y)-F^{-+}(x,y)\right),
\end{eqnarray}
%
where $\Theta(x_0-y_0)$ is the step function of the time difference.  The
rules for the co-contour functions $F_{--}$ etc. follow from Eq. (\ref{Fij}).

For such two point functions complex conjugation implies
%
\begin{eqnarray}\label{ComplexConjugate}
\left(\ii F^{-+}(x,y)\right)^*&=&\ii F^{-+}(y,x)
\quad\Rightarrow\quad \ii F^{-+}(X,p)=\mbox{real},\nonumber\\
\left(\ii F^{+-}(x,y)\right)^*&=&\ii F^{+-}(y,x)
\quad\Rightarrow\quad \ii F^{+-}(X,p)=\mbox{real},\nonumber\\
\left(\ii F^{--}(x,y)\right)^*&=&\ii F^{++}(y,x)
\quad\Rightarrow\quad \left(\ii F^{--}(X,p)\right)^*=\ii F^{++}(X,p),
\nonumber\\
\left(F^R(x,y)\right)^*&=&F^A(y,x)
\quad\hspace*{3.5mm}\Rightarrow\quad \left(F^R(X,p)\right)^*=F^A(X,p),
\end{eqnarray}
%
where the right parts specify the corresponding properties in the Wigner
representation. Diagrammatically these rules imply the simultaneous swapping
of all $+$ vertices into $-$ vertices and vice versa together with reversing
the line arrow-sense of all propagator lines in the diagram.

Contrary to the common case (\ref{Fxy}), the symmetrized contour convolution 
%
\begin{eqnarray}\label{S=(FG)}
E(x)=\oint\di z \left[F(x,z)G(z,x) + G(x,z)F(z,x)\right]
\end{eqnarray}
%
is continuous, when two contour coordinates become identical. 
This can be easily checked, proceeding from relations
(\ref{Fretarded}) for $F$ and $G$ functions. Moreover, for this 
symmetrized convolution with two coincident points 
we obtain a very simple expression in the Wigner representation, if
all gradient corrections to the convolution are neglected (so-called
local approximation), 
%
\begin{eqnarray}\label{S=(FG)loc}
E^{\scr{loc}}(X)=\int \dpi{p} \left[F^{--}(X,p)G^{--}(X,p) - 
F(X,p)^{++}G^{++}(X,p)\right].
\end{eqnarray}
%
In particular, this form is applicable to the potential
(\ref{eps-potk-K-B}) and derivative (\ref{var-tens-pN}) energy
densities.

\section{Derivative Coupling}\label{Der.-Coupl.}

To be specific, 
we consider a multicomponent system with different constituents $a$
described by nonrelativistic fermionic and relativistic scalar bosonic field
operators, 
summarized as $\Ph=\{\Pa(x)\}$. The free Lagrangians of these fields are 
%
\begin{eqnarray}
\label{L0}
\Lgh^0_a=\left\{
\begin{array}{ll}
\frac{1}{2}
\displaystyle\left(
\ii \Pad\partial_t \Pa  
-\ii \partial_t \Pad\Pa 
- \frac{1}{m_a} \nabla \Pad \nabla \Pa\right) 
\hspace*{-4mm}\quad& {\mbox{nonrel. fermions}}\\[2mm] 
\frac{1}{2}\displaystyle\vphantom{\frac{1}{m_a}}
 \left(\partial_\mu\Pa \cdot \partial^\mu\Pa 
-  m^2_a \Pa^2\right) 
\quad&{\mbox{neutral rel. bosons}}\\[2mm]
\displaystyle\vphantom{\frac{1}{m_a}}
\partial_\mu\Pad  \partial^\mu\Pa 
- m^2_a \Pad\Pa 
\quad&{\mbox{charged rel. bosons}}\\ 
\end{array}\hspace*{-1cm}\right.
\end{eqnarray}
%
We assume that these fields interact via linear derivative coupling,
such that the interaction Lagrangian does not only depend on these
fields but also on their derivatives
$\Lint=\Lint\{\Pa,\Pad,\partial^\mu\Pa,\partial^\mu\Pad\}$.  The
variational principle of stationary action determines Euler--Lagrange
equations of motion for the field operators
%
\begin{eqnarray}
\label{eqmotionL}
\partial_\mu\frac{\partial \Lgh^0}{\partial\left(\partial_\mu\Pad\right)}
-\frac{\partial \Lgh^0}{\partial\left(\Pad\right)}&=&
\frac{\partial \Lint}{\partial\left(\Pad\right)} - 
\partial_\mu\frac{\partial \Lint}{\partial\left(\partial_\mu\Pad\right)}
=: \frac{\delta \Lint}{\delta\Pad(x)}
\end{eqnarray}
%
and the corresponding adjoint equation, where 
the ``variational'' $\delta$-derivative is defined as 
%
\begin{eqnarray}\label{delta-derivative}
  \frac{\delta}{\delta f(x)}\dots
  :=\frac{\partial}{\partial f(x)}\dots
  -\partial_\mu\left(\frac{\partial}
    {\partial(\partial_\mu f(x))}\dots\right). 
\end{eqnarray}
%
This is the key definition, which allows to recast the local-coupling
formulas to the derivative coupling case. In fact, the ``variational''
$\delta$-derivative specifies the {\em full} derivative over $f(x)$,
implying that all derivatives acting on $f(x)$ in the action should be
redirected to other terms by means of partial integration before
taking variational derivatives of $f(x)$.

The equations of motion can also be written in the differential form
%
\begin{eqnarray}
\label{eqmotion}
\Gr_0^{-1}(x) \Pa(x)&=& - \ja(x)\equiv - \frac{\delta \Lint}{\delta\Pad(x)}
\end{eqnarray}
%
and similarly for the corresponding adjoint equation.
The $\ja(x)$  operator is 
the local source current of field $a$ with mass $m_a$, and
$\Gr_0^{-1}(x)$ is the free evolution operator\footnote{Note that the
first line in (\ref{S-def}) is not the nonrelativistic limit of the
second one. We have already divided the second line by $2m_a$, to take
into account different normalizations
of relativistic and nonrelativistic wave functions.}
%
\begin{eqnarray}
\label{S-def}             
\Gr_0^{-1}(x)=\left\{
\begin{array}{ll}
\displaystyle -\partial_\mu\partial^\mu -m_a^2
\quad&\mbox{\rm for rel. bosons}\\
\displaystyle\ii\partial_t+\frac{1}{2m_a}{\partial_{\vecs x}^2}
\quad&\mbox{\rm for nonrel. particles}
\end{array}\right.
\end{eqnarray}
%
with free propagator $\Gr_0(y,x)$ as resolvent, cf. Eq. (\ref{G0}).

Invariances of the Lagrangian provide a set of conservation laws, the
most prominent of which are those for the energy--momentum and certain
currents. In addition to the standard canonical energy--momentum tensor
\cite{Itz80}, different representations of this tensor  
have been considered \cite{Belinfante,CalColJac}. 
Using the Euler--Lagrange equations of motion and the
definition of the source current (\ref{eqmotion}), one can show that the
following form also defines a conserving energy--momentum tensor
%
\begin{eqnarray}
\label{E-M-new-tensor}
\nonumber
\medhat{\Theta}^{\mu\nu}(x)&=&
-\frac{1}{2}\left[
\suma\Fnb
\left(\partial_x^\nu-\partial_y^\nu\right)
\left(
\frac{\partial \Lgh^{0}(x)}{\partial\left(\partial_\mu\Pa\right)}
\Pa(y)
-
\Pad(x)
\frac{\partial \Lgh^{0}(y)}{\partial\left(\partial_\mu\Pad\right)}
\right)
\right]_{x=y}
\nonumber
\\
&+&
g^{\mu\nu}
\left(
{\widehat{\cal E}}^{\scr{int}}(x)-
{\widehat{\cal E}}^{\scr{pot}}(x)
\right)
+{\widehat{\cal E}}_{\scr{(der.)}}^{\mu\nu}(x).
\end{eqnarray}
%
Here we have introduced the operators of the
interaction-energy density ${\widehat{\cal E}}^{\scr{int}}$ 
and the potential-energy density ${\widehat{\cal E}}^{\scr{pot}}$
%
\begin{eqnarray}
\label{eps}
{\widehat{\cal E}}^{\scr{int}}(x) = - \Lint(x), 
\end{eqnarray}
%
%
\begin{eqnarray}
\label{eps-pot}
{\widehat{\cal E}}^{\scr{pot}}(x)
=-\frac{1}{2}\suma\Fnb
\left(
\jad(x) \Pa(x) + \Pad(x) \ja(x) 
\right). 
\end{eqnarray}
%
Furthermore, we have singled out the contribution
%
\begin{eqnarray} 
\label{E-M-der-tensor}
{\widehat{\cal E}}_{\scr{(der.)}}^{\mu\nu}(x) =
\suma\Fnb
\left(
\frac{\partial \Lint}{\partial\left(\partial_\mu\Pa\right)}
\cdot \partial^\nu\Pa
+ 
\partial^\nu \Pad \cdot 
\frac{\partial \Lint}{\partial\left(\partial_\mu\Pad\right)}
\right)
\end{eqnarray}
%
arising in the case of derivative coupling.  Here and below the case
of neutral bosons results from equating $\Pa=\Pad$ in all the
formulas. Proper counting is assured by the extra $\Fnb$ factor which
takes the value $1/2$ for neutral boson (real fields) and 1 for
complex fields.

If the Lagrangian is invariant under some global transformation of charged
fields (with the charges $e_a$), e.g.,
%
\begin{equation}
\label{c-global-tr.} 
\Pa(x)\Rightarrow e^{-\ii e_a\Lambda}\Pa(x);\quad\quad
\Pad(x)\Rightarrow e^{\ii e_a\Lambda}\Pad(x), 
\end{equation}
%
there exists a Noether current defined as \cite{Itz80} 
%
\begin{equation}
\label{c-new-current} 
\medhat{j}^{\mu}
= 
-\ii \suma
e_a \left(
\frac{\partial \Lgh}{\partial\left(\partial_\mu\Pa\right)}\Pa
- 
\Pad \frac{\partial \Lgh}{\partial\left(\partial_\mu\Pad\right)}
\right) =\medhat{j}^{\mu}_{\scr{(conv.)}}+
\medhat{j}^{\mu}_{\scr{(der.)}}. 
\end{equation}
%
which is conserved, i.e. 
$\partial_{\mu} \medhat{j}^{\mu} = 0$. 
Here, we have decomposed it into two terms: the conventional one
%
\begin{equation}
\label{c-current-conv} 
\medhat{j}^{\mu}_{\scr{(conv.)}}=
-\ii \suma
e_a \left(
\frac{\partial \Lgh^{0}}{\partial\left(\partial_\mu\Pa\right)}\Pa
- 
\Pad \frac{\partial \Lgh^{0}}{\partial\left(\partial_\mu\Pad\right)}
\right),
\end{equation}
%
which is associated with the free Lagrangian, and the derivative term 
%
\begin{equation}
\label{c-current-der} 
\medhat{j}^{\mu}_{\scr{(der.)}}=
-\ii \suma
e_a \left(
\frac{\partial \Lint}{\partial\left(\partial_\mu\Pa\right)}\Pa
- 
\Pad \frac{\partial \Lint}{\partial\left(\partial_\mu\Pad\right)}
\right)
\end{equation}
%
which is non-zero only for derivative coupling. 

To define the $\Phi$-functional for the case under consideration, all the
steps described in ref. \cite{IKV} should be repeated. Then we arrive to the
$\Phi$-functional that depends also on the gradients of mean fields
($\partial_\mu\Pba$ and $\partial_\mu\Pba^*$) and Green functions
($\partial_x^\mu\Gr(x,y)$ and $\partial_y^\mu\Gr(x,y)$) rather than on their
values only. The variational rules of this functional formally look similarly
to those in ref. \cite{IKV}
%
\begin{eqnarray}\label{varphdl}
    \ii\jba(x)&=&\frac{\delta \ii\Phi}{\delta \Pba^{\ast} (x)},\\
  \label{varphdl1}
  -\ii \Se_a(x,y)&=&\mpa\frac{\delta \ii \Phi}{\delta \ii\Gr_a(y,x)}
\times
\left\{
\begin{array}{ll}
2\quad&\mbox{for real fields}\\
1\quad&\mbox{for complex fields}
\end{array}\right.\\
\label{varphdl2}
- {\cal E}^{\scr{int}}(x)&=& 
\frac{\delta\ii \Phi}{\delta \ii\lambda (x)}, 
\end{eqnarray}
%
but should be understood in terms of the variational $\delta$-derivative of
Eq. (\ref{delta-derivative}) for one-point functions (like $\Pba(x)$ and
$\lambda(x)$) and its generalization 
%
\begin{eqnarray}\label{delta-derivative2}
  \frac{\delta\ii\Phi}{\delta\ii\Gr(y,x)}
  :=\frac{\delta_0\ii\Phi}{\delta_0\ii\Gr(y,x)}
  &-&\partial_x^\mu\left(\frac{\delta_0\ii\Phi}
    {\delta_0(\partial_x^\mu \ii\Gr(y,x))}\right)
  -\partial_y^\mu\left(\frac{\delta_0\ii\Phi}
    {\delta_0(\partial_y^\mu \ii\Gr(y,x))}\right)\cr
  &+&\partial_x^\mu\partial_y^\nu\left(\frac{\delta_0\ii\Phi}
    {\delta_0(\partial_x^\mu\partial_y^\nu \ii\Gr(y,x))}\right)
\end{eqnarray}
%
to two-point functions. Here
$\delta_0/\delta_0\ii\Gr(y,x)$ means the conventional
variation over $\Gr(y,x)$, which does not touch $\partial_x^\mu\ii\Gr$,
$\partial_y^\mu\ii\Gr$ and $\partial_x^\mu\partial_y^\nu\ii\Gr$ 
terms in $\ii\Phi$. Similarly to the variational $\delta$-derivative of
Eq. (\ref{delta-derivative}), the $\delta$-variation of
Eq. (\ref{delta-derivative2}) means the {\em full} variation over
$\Gr(y,x)$, implying that all derivatives acting on $\Gr(y,x)$ in the
$\Phi$ functional should be redirected to other terms by means of
partial integration before taking variation over $\Gr(y,x)$. 
The factor $\lambda (x)$, 
appeared in Eq. (\ref{varphdl2}), is an auxiliary scaling parameter of the
coupling constant. In terms of the $\Phi$ functional the additional derivative
contributions to mean values of the energy--momentum tensor
(\ref{E-M-der-tensor}) and current (\ref{c-current-der}) take the following
form 
%
\begin{eqnarray}
\label{var-tens}
{\cal E}_{\scr{(der.)}}^{\mu\nu}
&=:&\left< \medhat{\cal E}_{\scr{(der.)}}^{\mu\nu}\right>
=\displaystyle 
\suma\left(\Fnb
\vphantom{\oint} 
\left[ 
\frac{\delta\Phi}{\delta(\partial_\mu\Pba(x))}\partial^\nu\Pba(x)
+
\frac{\delta\Phi}{\delta(\partial_\mu\Pba^*(x))}\partial^\nu\Pba^*(x)
\right]\right. 
\nonumber
\\
&+&\displaystyle
\left.
\oint\di z\left[
\frac{\delta\Phi}{\delta(\partial_\mu^x\ii\Gr_a(z,x))} \cdot
\partial^\nu_x\ii\Gr_a(z,x)
+
\partial^\nu_x\ii\Gr_a(x,z) \cdot
\frac{\delta\Phi}{\delta(\partial_\mu^x\ii\Gr_a(x,z))}\right]\right),  
\end{eqnarray}
%
%
\begin{eqnarray}
\label{var-vec}
j_{\scr{(der.)}}^\mu &=:&\left< \medhat{j}^{\mu}_{\scr{(der.)}}\right>
=\displaystyle -\ii\suma e_a 
\left(
\vphantom{\oint} 
\left[ 
\frac{\delta\Phi}{\delta(\partial_\mu\Pba(x))}\Pba(x)
-
\frac{\delta\Phi}{\delta(\partial_\mu\Pba^*(x))}\Pba^*(x)
\right]\right. 
\nonumber
\\
&+&\displaystyle 
\left.
\oint\di z\left[
\ii\Gr_a(x,z) \cdot \frac{\delta\Phi}{\delta(\partial_\mu^x\ii\Gr_a(x,z))}
-
\frac{\delta\Phi}{\delta(\partial_\mu^x\ii\Gr_a(z,x))} \cdot \ii\Gr_a(z,x)
\right]\right),  
\end{eqnarray}
%
while the rest terms of $\Theta^{\mu\nu}$ and $j^\mu$ retain the same form as
that for local coupling, cf. ref. \cite{IKV}. 
Here the variation is also understood in terms of Eq. (\ref{delta-derivative2})
to take account of the $\partial_x^\mu\partial_y^\nu \ii\Gr_a(y,x)$ dependence
of the $\Phi$ functional.

The next step to the kinetic description consists in gradient
expansion of Kadanoff--Baym equations and all the related
quantities. Expansion of the equations of motion up to the first
order in gradients implies that the conserving quantities and
self-energies, except for possible memory terms in the collision
integral, are required only up to zero order in gradients. These
zero-order quantities are determined by the local $\Phi$ functional,
where all gradient corrections are neglected. Since in the local
approximation $\partial_x^\mu\ii\Gr$,
$\partial_y^\nu\ii\Gr$ and $\partial_x^\mu\partial_y^\nu\ii\Gr$
transform into $-\ii q^\mu\ii\Gr(X,q)$,
$\ii q^\nu\ii\Gr(X,q)$ and $-\ii q^\mu \ii q^\nu\ii\Gr(X,q)$,
respectively, no partial integrations are needed for the
variations of Eqs. (\ref{varphdl1})--(\ref{varphdl2}). This means that 
conventional variation rules of Eq. (\ref{var-Phi-component}) still
hold in this case. At the same time, derivative contributions to
the conserving quantities, Eqs. (\ref{var-tens}) and (\ref{var-vec}),
involve only variations over derivatives of the Green functions and,
hence, should be carefully defined within the local approximation for
the particular application considered.

\section{Nonrelativistic Nuclear Matter}\label{Potent}

Currently, calculations of ground-state and low-temperature properties
of nuclear matter 
are performed within the $G$- or $T$-matrix approximations to the
self-energy \cite{Bozek99,Dickhoff,Dewulf}.  Based on realistic
nonrelativistic nucleon--nucleon potentials, they quantitatively reproduce
phenomenological properties of nuclear matter. However, already for the
ground state the resulting
chemical potential, i.e. the single particle separation energy, deviates from
the binding energy per particle violating the
Hugenholtz--van-Hove theorem. This is a manifestation of problems with
the thermodynamic consistency in these approximations,
which gets even worse at nonzero temperatures. This problem was
discussed in refs. \cite{Bozek99,Baldo90,Jong91}. A consistent way to
overcome this problem consists in using a self-consistent $T$-matrix
approximation \cite{Bozek99} based on the $\Phi$-derivable
approximation. 

Dynamic simulations of the nuclear matter are even more demanding to
the choice of approximation to the self-energy, because the
requirement of charge and energy--momentum conservations should be met
except for that of the thermodynamic consistency. Again all these
requirements are met provided the approximation is
$\Phi$-derivable. Since dynamic simulations are much more
complicated as compared to static ones, up to now they were performed
in a simpler approximation to the self-energy, i.e. the direct Born
approximation \cite{D2,Bozek97}, which provide qualitative description
of the dynamics. These simulations were based on the Kadanoff--Baym
equations without any gradient expansion. Here we would like to call
attention to the fact that the use of the gradient expansion in the KB
form (see subsect. \ref{KB}) would simplify these dynamic simulations
and at the same time preserve the pleasant features of exact
conservations and thermodynamic consistency. 

In view of reasonable level of complexity
feasible for current computing, we confine our consideration to the
full Born approximation to the $\Phi$ functional 
%
\unitlength=0.8mm
\begin{eqnarray}\label{V-phi}\cr
\ii\Phi^{\scr{HFB}} =\;
\underbrace{\frac{1}{2}
\parbox{28mm}{
\begin{fmfgraph*}(19,19)
\fmfpen{thick}
\fmftop{t}
\fmfbottom{b}
\fmfforce{(0.5w,0.7h)}{t}
\fmfforce{(0.5w,0.3h)}{b}
\fmf{boson,width=thin,label=$V$,label.side=left}{t,b}
\fmf{fermion,left=1.,tension=.4}{t,t}
\fmf{fermion,left=1.,tension=.4}{b,b}
\fmfdot{t,b}
\end{fmfgraph*}
}
\hspace*{-12mm} + \frac{1}{2} \;\;\;
\parbox{28mm}{
\begin{fmfgraph*}(19,19)
\fmfpen{thick}
\fmftop{t}
\fmfbottom{b}
\fmfforce{(0.5w,1.0h)}{t}
\fmfforce{(0.5w,0.0h)}{b}
\fmf{boson,width=thin,label=$V$,label.side=left}{t,b}
\fmf{fermion,left=1.,tension=.4}{t,b}
\fmf{fermion,left=1.,tension=.4}{b,t}
\fmfdot{t,b}
\end{fmfgraph*}
}\hspace*{-10mm}
\displaystyle
\vphantom{
\frac{\frac{\int}{\displaystyle\int}}{\displaystyle\frac{\int}{\displaystyle\int}}
}
}_{\displaystyle\Phi^{\scr{HF}}}
+ \underbrace{\frac{1}{4} \;\;\;
\parbox{28mm}{
\begin{fmfgraph}(19,19)
\fmfpen{thick}
\fmfleft{lt,lb}
\fmfright{rt,rb}
\fmfforce{(1.0w,1.0h)}{rt}
\fmfforce{(1.0w,0.0h)}{rb}
\fmfforce{(0.0w,1.0h)}{lt}
\fmfforce{(0.0w,0.0h)}{lb}
\fmf{boson,width=thin}{lt,lb}
\fmf{boson,width=thin}{rt,rb}
\fmf{fermion,left=0.3,tension=.4}{lt,rt}
\fmf{fermion,left=0.3,tension=.4}{rt,lt}
\fmf{fermion,left=0.3,tension=.4}{lb,rb}
\fmf{fermion,left=0.3,tension=.4}{rb,lb}
\fmfdot{lt,lb,rt,rb}
\end{fmfgraph}
}\hspace*{-10mm}
+ \frac{1}{4} \;\;\;
\parbox{28mm}{
\begin{fmfgraph}(19,19)
\fmfpen{thick}
\fmfleft{lt,lb}
\fmfright{rt,rb}
\fmfforce{(1.0w,1.0h)}{rt}
\fmfforce{(1.0w,0.0h)}{rb}
\fmfforce{(0.0w,1.0h)}{lt}
\fmfforce{(0.0w,0.0h)}{lb}
\fmfforce{(0.5w,0.5h)}{c}
\fmf{boson,width=thin}{lt,lb}
\fmf{boson,width=thin}{rt,rb}
\fmf{fermion,tension=.4}{lt,rt}
\fmf{plain,tension=.4}{rt,lb}
\fmf{phantom_arrow,tension=.4}{rt,c}
\fmf{fermion,tension=.4}{lb,rb}
\fmf{plain,tension=.4}{rb,lt}
\fmf{phantom_arrow,tension=.4}{rb,c}
\fmfdot{lt,lb,rt,rb}
\end{fmfgraph}
}\hspace*{-10mm}
\displaystyle
\vphantom{
\frac{\frac{\int}{\displaystyle\int}}{\displaystyle\frac{\int}{\displaystyle\int}}}
}_{\displaystyle\Phi^{\scr{Born}}}\hspace*{6mm} 
\\[3mm]
\nonumber
\end{eqnarray}
%
which includes the Hartree--Fock contribution $\Phi^{\scr{HF}}$ (the
first two diagrams in Eq. (\ref{V-phi})) and the true Born
contribution $\Phi^{\scr{Born}}$ (the last two diagrams). Here the
wave line symbolizes a nonlocal nucleon--nucleon potential 
$V(\mid\vec{x}_1 - \vec{x}_2\mid)$, or $V(\mid\vec{q}\mid)$
in the momentum representation. For simplicity, below we denote the
latter as $V(q)$, keeping in mind that in fact it does not depend on
neither $q_0$ nor direction of $\vec{q}$. 

Note that the $\Phi^{\scr{Born}}$ part gives rise to the self-energy
containing internal vertices. This implies that the corresponding
collision term involves non-local effects (see discussion in
ref. \cite{IKV99}). However, only ``spatial non-locality'' appears in the
collision term, while the memory in time is absent since
$V(\mid\vec{x}_1 - \vec{x}_2\mid)\delta(t_1 - t_2)$ is time local. 
According to the general consideration of ref. \cite{KIV01}, exact
conservations in the gradient approximation take place if all the
non-local terms are consistently taken into account up to first-order
gradients. Below we show that in the particular case of the
$\Phi^{\scr{HFB}}$ functional the exact conservations hold true even
if we neglect the ``spatial non-locality'' generated by
$\Phi^{\scr{HFB}}$. These exact conservations imply the
global rather than local conservation of the energy--momentum, which
is in fact natural for the case of instant interaction of finite range
considered here. 

Neglecting the gradient terms induced by the finite range of $V$, we
consider the $\Phi^{\scr{Born}}$ functional in the local
approximation, where all Green functions in the Wigner representation
are taken at the same centroid coordinate $X$. Alongside in some
variational expressions we use an $X$-dependent local $\Phi$
functional, $\Phi(X)$, where the last spatial
integration is omitted, i.e.
%
\begin{eqnarray}\label{Phi(X)}
\Phi=\int\di X \Phi(X)
\end{eqnarray}
%
The $\Phi^{\scr{HFB}}$ of
Eq. (\ref{V-phi}) gives rise to the following local collision term 
%
\begin{eqnarray}
\label{HBF-C}
C^{\scr{HFB-loc}}
&=& \int \dpi{p_2} \dpi{p_3} \dpi{p_4}
R^{\scr{HFB}}
\nonumber\\
&\times&
\left(\Ft_1\Ft_2\F_3\F_4 - \F_1\F_2\Ft_3\Ft_4\right)
\delta^4 (p_1+p_2-p_3-p_4),  
\\
\label{gain-V}
R^{\scr{HFB}} &=& \frac{(2\pi)^4}{2}  
\left[
V(p_1-p_3) + V(p_1-p_4) 
\right]^2, 
\end{eqnarray}
%
where $\F_1=\F (X,p_1)$, etc., cf. Eq. (\ref{Coll-var}).

\subsection{Charge Conservation}\label{Nonrel.Charge}

In $\Phi^{\scr{HFB}}$ the $\Gr^{--}$ and $\Gr^{++}$ Green functions
are encountered only in different $+-$ 
$\Phi$-diagrams, and hence we can vary $\Gr^{--}$ and $\Gr^{++}$
independently. Therefore, $\Phi^{\scr{HFB}}$ is invariant under the following
transformation 
%
\begin{eqnarray}\label{xi-Btr}
\Gr^{--}(X,p)\,\, \Rightarrow\,\, \Gr^{--}(X,p+\xi(X)), \quad
\Gr^{++}(X,p)\,\, \Rightarrow\,\, \Gr^{++}(X,p-\xi(X))
\end{eqnarray}
%
with $\F$, $\Ft$ and $V$ kept unchanged. Here $\xi(X)$ is an
arbitrary function. If $|\xi(X)|\ll 1$, transformation (\ref{xi-Btr}) reads 
%
\begin{eqnarray}\label{xi-Btr-inf}
\delta \Gr^{--} = \xi_\mu(X) \frac{\partial\Gr^{--}}{\partial p_\mu}, \quad
\delta \Gr^{++} =-\xi_\mu(X) \frac{\partial\Gr^{++}}{\partial p_\mu}.
\end{eqnarray}
%
Performing variation of $\Phi^{\scr{Born}}$ under the transformation
(\ref{xi-Btr-inf}) within the canonical variation rules
(\ref{var-Phi-component}), we arrive at 
%
\begin{eqnarray}
\label{NB-cons} 
\ii\delta\Phi^{\scr{loc}}&=&
\int \di X \xi_\mu(X)\Tr\int \dpi{p}
\left(
\ii\Se_{--}\frac{\partial\ii\Gr^{--}}{\partial p_\mu} -
\ii\Se_{++}\frac{\partial\ii\Gr^{++}}{\partial p_\mu}
\right)  
\nonumber\\
&=& 
2\ii\int \di X \xi_\mu(X)\Tr\int \dpi{p}
\left(
\Ldt\frac{\partial\Re\Gr^R}{\partial p_\mu} +
\Re\Se^R\frac{\partial\F}{\partial p_\mu}
\right)  
=0.   
\end{eqnarray}
%
Here we have used that the  integral 
%
\begin{eqnarray}
\label{NBR-cons} 
\Tr\int \dpi{p}
\left(
\Gm\frac{\partial\Re\Gr^R}{\partial p_\mu} +
\Re\Se^R\frac{\partial\A}{\partial p_\mu}
\right)
=
-\frac{1}{2}\Im \; \Tr\int \dpi{p}
\Se^R\frac{\partial\Gr^R}{\partial p_\mu}
=0.   
\end{eqnarray}
%
equals zero due to analyticity of $\Gr^R$ and $\Se^R$. Thus, we obtain 
the relation 
%
\begin{eqnarray}
\label{NBB-cons} 
\Tr\int \dpi{p}
\left(
\Ldt\frac{\partial\Re\Gr^R}{\partial p_\mu} +
\Re\Se^R\frac{\partial\F}{\partial p_\mu}
\right)=0,    
\end{eqnarray}
%
{\em which guarantees the Noether current conservation},
cf. Eq. (\ref{invarJk}) with $j^{\mu}_{\scr{(der)}}=0$.

\subsection{Energy--Momentum Conservation}\label{Nonrel.Energy}

In order to construct the conservation laws related to space--time
homogeneity we perform the following transformation  
%
\begin{eqnarray}\label{xiX-Btr}
\Gr^{--}(X,p)\,\, \Rightarrow\,\, 
\Gr^{--}\left(X+\xi(X), p\right), 
\quad
\Gr^{++}(X,p)\,\, \Rightarrow\,\, 
\Gr^{++}\left(X-\xi(X), p\right)
\end{eqnarray}
%
with $\F$, $\Ft$ and $V$ 
kept unchanged. This transformation
only acts on $\Phi^{\scr{HFB}-}$ and $\Phi^{\scr{HFB}+}$, i.e. those
involving only "$-$" or "$+$" vertices, respectively, 
%
\begin{eqnarray}\label{xiX-calc}
\delta\Phi^{\scr{HFB}}= 
\int \di X \xi_\mu(X) \partial^\mu 
\left(\Phi^{\scr{HFB}}(X^{-})-\Phi^{\scr{HFB}}(X^{+})\right)
\end{eqnarray}
%
where $\Phi^{\scr{HFB}}(X^{i})$ are understood in the sense of
(\ref{Phi(X)}). Note that 
%
\begin{eqnarray}\label{Phi-Phi}
&&\ii\left(\Phi^{\scr{HFB}}(X^{-})-\Phi^{\scr{HFB}}(X^{+})\right)=
\ii\Tr\int \dpi{p}
\Se^{\scr{HF}}\F
\nonumber\\
&+&\frac{1}{2}\ii\Tr\int \dpi{p}
\left(
\Ldt\Re\Gr^R + (\Re\Se^R-\Se^{\scr{HF}})\F
\right)  
\nonumber\\
&-& 
\underbrace{\Tr\int \dpi{p}
\left(
\Gm\Re\Gr^R + (\Re\Se^R-\Se^{\scr{HF}})\A
\right)}_{=0},   
\end{eqnarray}
%
where the last integral is again zero due to analyticity, similarly to
(\ref{NBR-cons}). Here the
first term on the r.h.s. results from the first two (Hartree--Fock) diagrams
in Eq. (\ref{V-phi}), while the last two integrals follow from the last two 
(Born) diagrams. Alternatively, we can perform variation of
$\Phi^{\scr{Born}}$ applying the canonical variation rules
(\ref{var-Phi-component}) 
%
\begin{eqnarray}\label{xiX-dPhi}
\ii\delta\Phi^{\scr{HFB}}&=& 
\int \di X \xi_\mu(X) \Tr\int \dpi{p}
\left(
\ii\Se_{--}\frac{\partial\ii\Gr^{--}}{\partial X_\mu} -
\ii\Se_{++}\frac{\partial\ii\Gr^{++}}{\partial X_\mu}
\right)  
\nonumber\\
&=& 
2\ii\int \di X \xi_\mu(X)\Tr\int \dpi{p}
\left(
\Ldt\frac{\partial\Re\Gr^R}{\partial X_\mu} +
\Re\Se^R\frac{\partial\F}{\partial X_\mu}
\right)  
\nonumber\\
&&- 
\ii\underbrace{\int \di X \xi_\mu(X)\Tr\int \dpi{p}
\left(
\Gm\frac{\partial\Re\Gr^R}{\partial X_\mu} +
\Re\Se^R\frac{\partial\A}{\partial X_\mu}
\right)}_{=0}, 
\end{eqnarray}
%
where the last integral is again zero due to analyticity. 
Therefore, we arrive at the important identity 
%
\begin{eqnarray}\label{X-dPhi}
&&\Tr\int \dpi{p}
\left(
\Ldt\frac{\partial\Re\Gr^R}{\partial X_\mu} +
\Re\Se^R\frac{\partial\F}{\partial X_\mu}
\right)  
\nonumber\\
&=&\partial^\mu 
\Tr\int \dpi{p}\left[
\frac{1}{2}\Se^{\scr{HF}}\F
+
\frac{1}{4}
\left(
\Ldt\Re\Gr^R + (\Re\Se^R-\Se^{\scr{HF}})\F
\right)\right].  
\end{eqnarray}
%

Next we investigate the following transformation
%
\begin{eqnarray}\label{xiLP-Btr}
&&
\Gr^{--}(X,p)\,\, \Rightarrow\,\, 
\Gr^{--}\left(X, \Lambda_{\mu\nu}(X)p^\nu\right), 
\quad
\Gr^{++}(X,p)\,\, \Rightarrow\,\, 
\Gr^{++}\left(X, \Lambda_{\mu\nu}^{-1}(X)p^\nu\right)
\\
\label{xiV-Btr}
&&V^-(q)\,\, \Rightarrow\,\,  V^-(\Lambda_{\mu\nu}(X)q^\nu),
\quad
V^+(q)\,\, \Rightarrow\,\,  V^+(\Lambda_{\mu\nu}^{-1}(X)q^\nu)
\end{eqnarray}
%
for the entire $\Phi^{\scr{HFB}}$, while $\F$, $\Ft$ are kept
unchanged. For the Hartree--Fock part $\Phi^{\scr{HF}}$ one finds
%
\begin{eqnarray}\label{xiLP-calc-HF}
&&
\tilde{p}_\mu = \Lambda_{\mu\nu} p^\nu,\quad
\tilde{p}'_\mu = \Lambda_{\mu\nu} p'^\nu \quad
\Rightarrow\quad
\di^4 p\;\di^4 p'= 
(\det\Lambda)^{-2} \; 
\di^4 \tilde{p}\;\di^4 \tilde{p}',
\\&&
\delta\Phi^{\scr{HF}-} = \int \di X
\left[(\det\Lambda)^{-2}-1\right] \Phi^{\scr{HF}}(X^{-}),
\\&&
\delta\Phi^{\scr{HF}+} = \int \di X
\left[(\det\Lambda)^{2}-1\right] \Phi^{\scr{HF}}(X^{+}),
\end{eqnarray}
%
where again $\Phi^{\scr{HF}-}$ and $\Phi^{\scr{HF}+}$ are the
$\Phi^{\scr{HF}}$ diagrams involving only "$-$" or "$+$" vertices,
respectively.  In general, an arbitrary diagram
$\Phi^{n_G,n_\lambda(-)}$, consisting of $n_G$  Green functions,
$n_\lambda$  of "$-$" interactions and no "$+$" interactions,
transforms as
%
\begin{eqnarray}\label{var-Phi-scaling-}
\delta\Phi^{n_G,n_\lambda(-)} = \int \di X
\left[(\det\Lambda)^{-(n_G-n_\lambda+1)}-1\right] \Phi^{n_G,n_\lambda(-)}(X), 
\end{eqnarray}
%
since the change of each momentum integration gives $(\det\Lambda)^{-1}$
(cf. Eq. (\ref{xiLP-calc-HF})), while the transformation of the
$\delta(p)$-function at each 
vertex (apart from one vertex due to global momentum conservation of
the diagram) produces $\det\Lambda$
%
\begin{eqnarray}\label{xiLP-delt(p)}
\delta^4(p+p'-q)=
\det\Lambda \; \delta^4(\tilde{p}+\tilde{p}'-\tilde{q}). 
\end{eqnarray}
%
Similarly,  
%
\begin{eqnarray}\label{var-Phi-scaling+}
\delta\Phi^{n_G,n_\lambda(+)} = \int \di X
\left[(\det\Lambda)^{n_G-n_\lambda+1}-1\right] \Phi^{n_G,n_\lambda(+)}(X). 
\end{eqnarray}
%
According to these rules the "$-$" and "$+$" diagrams of the second order in
the interaction, $\Phi^{\scr{Born}-}$ and $\Phi^{\scr{Born}+}$, are
transformed as follows 
%
\begin{eqnarray}\label{xiLP-calc-}
&&
\delta\Phi^{\scr{Born}-} = \int \di X
\left[(\det\Lambda)^{-3}-1\right] \Phi^{\scr{Born}--}(X),
\\ 
\label{xiLP-calc+}
&&
\delta\Phi^{\scr{Born}+} = \int \di X
\left[(\det\Lambda)^3-1\right] \Phi^{\scr{Born}++}(X).
\end{eqnarray}
%
If the $\Lambda$-transformation is infinitesimal,
$\Lambda_{\mu\nu}(X)=1+\omega_{\mu\nu}(X)$ with $\mid\omega_{\mu\nu}\mid\ll 1$
and $\det\Lambda=1+\Tr\omega$, $\det(\Lambda^{-1})=1-\Tr\omega$, we obtain 
%
\begin{eqnarray}\label{xiLP-calc}
\delta\left(\Phi^{\scr{HFB}-}+\Phi^{\scr{HFB}+}\right)
=
&-& \int \di X 2\Tr\omega
\left(\Phi^{\scr{HF}-}(X)-\Phi^{\scr{HF}+}(X)\right)
\cr
&-& \int \di X 3\Tr\omega
\left(\Phi^{\scr{Born}-}(X)-\Phi^{\scr{Born}+}(X)\right),
\end{eqnarray}
%
cf. Eq. (\ref{Phi-Phi}). The $\Phi^{\scr{Born}-+}$ and 
$\Phi^{\scr{Born}+-}$ components, i.e. those containing both "$-$" and
"$+$" vertices, are modified by only the $V$ 
transformation. Moreover, this transformation leaves them invariant
%
\begin{eqnarray}\label{xiLP-dPhi_mp}
\hspace*{-8mm}
&&\ii\delta\Phi^{\scr{Born}-+}
=\int \di X \omega_{\mu\nu} \Tr\int \dpi{p} p^\nu
\left(
\frac{\delta\ii\Phi^{\scr{Born}-+}(X)}{\delta\ii V^-}
\frac{\partial\ii V^-}{\partial p_\mu}
-\frac{\delta\ii\Phi^{\scr{Born}-+}(X)}{\delta\ii V^+}
\frac{\partial\ii V^+}{\partial p_\mu}
\right)  
\nonumber\\
\hspace*{-8mm}
&=&
\int \di X \omega_{\mu\nu} \Tr\int \dpi{p} p^\nu
\left(
\frac{\delta\ii\Phi^{\scr{Born}-+}(X)}{\delta V^-}
V^-
-\frac{\delta\ii\Phi^{\scr{Born}-+}(X)}{\delta V^+}
V^+
\right) 
\frac{1}{V}\frac{\partial V}{\partial p_\mu} =0, 
\end{eqnarray}
%
since $\Phi^{\scr{Born}-+}$ is symmetric with respect to $V^-$ and $V^+$. 
Thus, 
%
\begin{eqnarray}\label{xiLP-dPhi1}
\hspace*{-8mm}
\delta\ii\Phi^{\scr{HFB}}
=&-& \int \di X (\Tr\omega)
\ii\Tr\int \dpi{p}\left[
2\Se^{\scr{HF}}\F
\right.\nonumber\\&-&
\frac{3}{2}
\left.
\left(
\Ldt\Re\Gr^R + (\Re\Se^R-\Se^{\scr{HF}})\F
\right)\right].   
\end{eqnarray}
%

Alternatively, we can perform variation of
$\Phi^{\scr{HFB}}$ applying the canonical variation rules
(\ref{var-Phi-component}) 
%
\begin{eqnarray}\label{xiLP-dPhi}
\hspace*{-8mm}
\ii\delta\Phi^{\scr{HFB}}&=& 
\int \di X \omega_{\mu\nu} \left[\Tr\int \dpi{p} p^\nu
\left(
\ii\Se_{--}\frac{\partial\ii\Gr^{--}}{\partial p_\mu} -
\ii\Se_{++}\frac{\partial\ii\Gr^{++}}{\partial p_\mu}
\right) + 2\ii Q^{\mu\nu}(X) \right]
\nonumber\\
\hspace*{-8mm}
&=& 
2\ii\int \di X \omega_{\mu\nu}\left[\Tr\int \dpi{p}p^\nu
\left(
\Ldt\frac{\partial\Re\Gr^R}{\partial p_\mu} +
\Re\Se^R \frac{\partial\F}{\partial p_\mu}
\right) + Q^{\mu\nu}(X) \right]  
\nonumber\\
\hspace*{-8mm}
&&-\underbrace{ 
\ii\int \di X \omega_{\mu\nu}\Tr\int \dpi{p}p^\nu
\left(
\Gm\frac{\partial\Re\Gr^R}{\partial p_\mu} +
\Re\Se^R \frac{\partial\A}{\partial p_\mu}
\right)}_{=0},  
\end{eqnarray}
%
where the last integral again equals zero due to analyticity. 
Here we have introduced the quantity 
%
\begin{eqnarray}\label{Pi}
2\ii Q^{\mu\nu}(X)=
\Tr\int \dpi{p} p^\nu
\left(
\frac{\delta\ii\Phi^{\scr{HFB}-}(X)}{\delta\ii V^-}
\frac{\partial\ii V^-}{\partial p_\mu}
-\frac{\delta\ii\Phi^{\scr{HFB}+}(X)}{\delta\ii V^+}
\frac{\partial\ii V^+}{\partial p_\mu}
\right)  
\end{eqnarray}
%
arising from variation over $V$. All we have to know about this
quantity is that it is $Q^{\mu\nu}=0$ when $\mu=0$ and/or
$\nu=0$. This property results from $p_0$-independence of
$V(\mid\vec{p}\mid)$. In particular this property yields 
%
\begin{eqnarray}\label{xiLP-dV}
\int \di^3 X \partial_\mu Q^{\mu\nu}(X) =0, 
\end{eqnarray}
%
where the $X$-integration runs only over space. 

Hence, comparing Eq. (\ref{xiLP-dPhi}) to Eq. (\ref{xiLP-dPhi1}), we
arrive at another important identity  
%
\begin{eqnarray}\label{LP-dPhi}
\hspace*{-8mm}
&& 
\Tr\int \dpi{p}p^\nu
\left(
\Ldt\frac{\partial\Re\Gr^R}{\partial p_\mu} +
\Re\Se^R\frac{\partial\F}{\partial p_\mu}
\right) + Q^{\mu\nu}(X)  
\nonumber\\
&=& 
-g^{\mu\nu}\Tr\int \dpi{p} \Se^{\scr{HF}}\F
-g^{\mu\nu}\frac{3}{4}\Tr\int \dpi{p}\left(
\Ldt\Re\Gr^R + (\Re\Se^R-\Se^{\scr{HF}})\F
\right).   
\end{eqnarray}
%

We turn now to the r.h.s. of the consistency relation for
energy--momentum conservation (\ref{epsilon-invk})
%
\begin{eqnarray}
\label{K}
K^{\nu} =
\mbox{Tr}
\int \di^3 X
\frac{p^\nu \di^4 p}{(2\pi )^4}
\left[
\Pbr{\Re\Se^R,\F} 
- 
\Pbr{\Re\Gr^R,\Ldt} 
\right] 
\end{eqnarray}
%
integrated over space, which is suitable for the global conservation. 
In this expression the local collision term (\ref{HBF-C})
drops out according to Eq. (\ref{C-conser}).   
It can be transformed by means of the identity
%
\begin{equation}
\label{f[]dp} 
\int \dpi{p} 
p^\nu \Pbr{\varphi,f} = 
\int \dpi{p}  
\left[ \partial^{\mu} 
\left(
p^\nu f \frac{\partial \varphi}{\partial p^{\mu}} 
\right) + 
 f \partial^{\nu}\varphi \right],  
\end{equation}
%
where $\varphi$ and $f$ are arbitrary functions, with the result
%
\begin{eqnarray}
\label{K1}
K^{\nu} &=&
-\Tr\int \di^3 X \partial^\mu 
\int 
\frac{\di^4 p}{(2\pi )^4}
p^\nu \left(
\Re\Se^R  \frac{\partial\F}{\partial p^\mu} 
+ 
\Ldt \frac{\partial\Re\Gr^R}{\partial p^\mu}  
\right) 
\cr 
&&- 
\Tr
\int\di^3 X \frac{\di^4 p}{(2\pi )^4}
\left(
\Re\Se^R \partial^\nu \F + \Ldt \partial^\nu \Re\Gr^R
\right).   
\end{eqnarray}
%
Now, applying identities (\ref{X-dPhi}), (\ref{xiLP-dV}) and
(\ref{LP-dPhi}) to the r.h.s. of Eq. (\ref{K1}), we obtain 
%
\begin{eqnarray}
\label{K2}
K^{\nu} &=&
\int \di^3 X \partial_\mu g^{\mu\nu}
\Tr\int \dpi{p} 
\frac{1}{2}
\left(
\Ldt\Re\Gr^R + \Re\Se^R\F
\right)   
\end{eqnarray}
%
which is precisely needed for the global conservation of the
Noether energy--momentum 
%
\begin{eqnarray}
\label{glob-cons} 
\frac{\partial}{\partial T} \int \di^3 X \Theta^{0\nu}(X) = 0,  
\end{eqnarray}
%
since for the case under consideration ${\cal
  E}^{\scr{int}}=\frac{1}{2}{\cal E}^{\scr{pot}}$,
cf. Eq. (\ref{int-spec}).

\section{Nucleon--Pion System}\label{Deriv}

For the discussion of the physical aspects of the nucleon--pion problem
we refer to ref. \cite{IKHV00}. Here we would like to clarify some
technical details. We choose the nonrelativistic form of pion--nucleon
interaction \cite{Weise}
%
\begin{eqnarray}
\label{pi-N-int}
\Lint=g \Psd
\left[({\vec\sigma}\cdot{\vec\nabla})({\vec\tau}\cdot{\vec\Pb})\right]\Ps,  
\end{eqnarray}
%
where $\Ps$ and $\Pb$ are non-relativistic nucleon and Klein-Gordon
pion field operators, respectively. We accept a simple approximation
defined by the following $\Phi$ functional
%
\unitlength=0.8mm
\begin{eqnarray}\label{piN-phi}\cr
\ii\Phi^{\pi N} &=& \frac{1}{2}
\hspace*{3mm}
\parbox{28mm}{
\begin{fmfgraph*}(19,19)
\fmfpen{thick}
\fmfleft{l}
\fmfright{r}
\fmfforce{(1.0w,0.5h)}{r}
\fmfforce{(0.0w,0.5h)}{l}
\fmf{boson,label=$\pi$,label.side=left}{l,r}
\fmf{fermion,label=$N$,label.side=left,left=1.,tension=.4}{l,r}
\fmf{fermion,label=$N$,label.side=left,left=1.,tension=.4}{r,l}
\fmfdot{l,r}
\end{fmfgraph*}}
\hspace*{-10mm}
= \frac{1}{2}\Tr\int\di X \dpi{p_1}\dpi{p_2}\dpi{q}
\delta^4 (p_1-p_2+q) 
\nonumber\\[5mm]
&\times&
\ii\Gr^{ij}(X,p_1)\;
(-\ii g)(-\ii{\vec q}\cdot{\vec\sigma}){\vec\tau}\;
\ii\Delta_{ij}(X,q)\;
(-\ii g)(\ii{\vec q}\cdot{\vec\sigma}){\vec\tau}\;
\ii\Gr^{ji}(X,p_2),   
\end{eqnarray}
%
where $\Gr$ and $\Delta$ are the nucleon and pion Green functions,
respectively, and $\Tr$ runs over spin and isospin indices.  Here we
have assumed an isotopically symmetric system, where the pion Green
functions of all isotopic charges coincide.  Though this approximation
is evidently oversimplified to produce quantitative results, cf. ref.
\cite{IKHV00}, it is able to reproduce qualitative features of the
dynamics. Moreover, this approximation is at the edge of the present
computing abilities. The formal basics of the $\Phi$-functional
formalism are given in Append. \ref{Der.-Coupl.}.

The charge current, defined by Eqs. (\ref{c-new-currentk}) and
(\ref{c-new-current}), relates to the baryon number conservation and
hence is trivial from the point of view of the pion--nucleon
interaction. Indeed, to prove the baryon number conservation we should
perform transformation (\ref{xi-Btr}) with $\F_N$, $\Ft_N$ and
$\Delta^{ij}$ kept unchanged. The pion Green functions $\Delta_{ij}$
are not subjected to this transformation, since pions are neutral from
the point of view of baryonic charge. All the subsequent considerations
are completely identical to that of the Fock diagram (the second term
in Eq. (\ref{V-phi})) and lead to the same final result
(\ref{NBB-cons}), i.e. to the exact Noether current conservation. 

The energy--momentum conservation is more instructive in this respect.
Before proceeding to the conservation laws themselves we should define
the derivative contribution to the energy--momentum tensor
(\ref{var-tens}). In our case of vanishing mean fields the pion Green
function enters the $\Phi$-functional only doubly differentiated.
Therefore the expression (\ref{var-tens}) takes the form
%
\begin{eqnarray}
\label{var-tens-pi}
\hspace{-8mm}
{\cal E}_{\scr{(der.)}}^{\mu\nu}
= \frac{1}{2}
\oint\di z 
&&\left(
\frac{\delta_0\Phi}{\delta_0(\partial_\lambda^z\partial_\mu^x\ii\Delta(z,x))} 
\cdot
\partial_\lambda^z\partial^\nu_x\ii\Delta(z,x)
\right.
\cr 
&&+
\left.
\partial_\nu^x\partial^\lambda_z\ii\Delta(x,z) \cdot
\frac{\delta_0\Phi}{\delta_0(\partial_\mu^x\partial_\lambda^z\ii\Delta(x,z))}
\right),  
\end{eqnarray}
%
where $\delta_0$ is already the conventional variation. In the Wigner
representation with due regard for Eq. (\ref{S=(FG)loc}) it transforms into  
%
\begin{eqnarray}
\label{var-tens-pN}
{\cal E}_{\scr{(der.)}}^{\mu\nu}(X)
&=& -\Tr\int \dpi{p_1}\dpi{p_2}\dpi{q}
\delta^4 (p_1-p_2+q) g^2
\nonumber\\
&\times&
\left[
\ii\Gr^{--}(X,p_1)\;
(- q^\nu \sigma^\mu){\vec\tau}\;
\ii\Delta^{--}(X,q)\;
(\ii{\vec q}\cdot{\vec\sigma}){\vec\tau}\;
\ii\Gr^{--}(X,p_2)
\right.
\nonumber\\
&-&
\left.
\ii\Gr^{++}(X,p_1)\;
(- q^\nu \sigma^\mu){\vec\tau}\;
\ii\Delta^{++}(X,q)\;
(\ii{\vec q}\cdot{\vec\sigma}){\vec\tau}\;
\ii\Gr^{++}(X,p_2)
\right].  
\end{eqnarray}
%
Contrary to usual convention, here we use Greek indices $\mu$ and
$\nu$ for the components of 3-vectors in order to distinguish them
from the ``$+-$'' summation indices. 
The potential energy density is still determined by the standard
expression (\ref{eps-potk-K-B}) but in terms of the sum over nucleons
and pions. Within the approximation of Eq. (\ref{piN-phi}), 
${\cal E}^{\scr{pot}}$ can be alternatively expressed as
%
\begin{eqnarray}
\label{eps-potk-pN} 
{\cal E}^{\scr{pot}}(X)
= {\cal E}^{\scr{pot}}_N + {\cal E}^{\scr{pot}}_\pi 
= \frac{3}{2} [\Phi^{\pi N-}(X) - \Phi^{\pi N+}(X)], 
\end{eqnarray}
%
where $\Phi^{\pi N-}$ ($\Phi^{\pi N+}$) refers to the $\Phi^{\pi N}$
functional with removed integration over $\di X$ and all vertices
being of ``$-$'' (``$+$'') type. In view of relation (\ref{int-spec}), 
%
\begin{eqnarray}
\label{eps-int-pot-pN} 
{\cal E}^{\scr{int}}(X)-{\cal E}^{\scr{pot}}(X)
= - \frac{1}{2} [\Phi^{\pi N-}(X) - \Phi^{\pi N+}(X)].  
\end{eqnarray}
%

\subsection{Energy--Momentum Conservation}\label{Der.Energy}

We briefly repeat the steps proving the exact energy--momentum
conservation for the nonrelativistic nuclear matter
(Append. \ref{Nonrel.Energy}) with the emphasis on the specifics of the
derivative coupling.  

First the transformation of Eq. (\ref{xiX-Btr}) for the
nucleon Green functions together with the corresponding transformation
of the pion Green functions
%
\begin{eqnarray}\label{xiX-Btr-pi}
\Delta^{--}(X,p)\,\, \Rightarrow\,\, 
\Delta^{--}\left(X+\xi(X), p\right), 
\quad
\Delta^{++}(X,p)\,\, \Rightarrow\,\, 
\Delta^{++}\left(X-\xi(X), p\right), 
\end{eqnarray}
%
with $\F_\pi$ and $\Ft_\pi$ being kept unchanged has to be performed.
This transformation is unaffected by the derivative coupling, and in
the similar way as before we arrive at the identity
%
\begin{eqnarray}\label{X-dPhi-pi}
&&\Tr\int \dpi{p}
\left[\left(
\Ldt_N\frac{\partial\Re\Gr^R}{\partial X_\mu} +
\Re\Se^R\frac{\partial\F_N}{\partial X_\mu}
\right) +  
\frac{1}{2}
\left(
\Ldt_\pi\frac{\partial\Re\Delta^R}{\partial X_\mu} +
\Re\Pi^R\frac{\partial\F_\pi}{\partial X_\mu}
\right)\right]  
\nonumber\\
&=&-\partial^\mu 
\left({\cal E}^{\scr{int}}-{\cal E}^{\scr{pot}}\right). 
\end{eqnarray}
%
Here $\Se$ and $\Pi$ refer to nucleon and pion self-energies,
respectively, and subscripts $N$ or $\pi$ correspondingly attribute a
quantity to either nucleon or pion subsystems. The r.h.s. of this
identity is written with due regards to Eq. (\ref{eps-int-pot-pN}).

Let us now perform the transformation (\ref{xiLP-Btr}) for both
nucleon and pion Green functions, as well as $q$ factors encountered
in vertices of $\Phi^{\pi N}$. Then the variation of $\Phi^{\pi N}$
gives 
%
\begin{eqnarray}\label{xiLP-calc-pi}
\delta\Phi^{\pi N}
&=&
\int \di X  \omega_{\mu\nu}
\left[ -2g^{\mu\nu}\left(\Phi^{\pi N-}(X)-\Phi^{\pi N+}(X)\right)
+ 2{\cal E}_{\scr{(der.)}}^{\mu\nu}\right]
\cr
&=& \int \di X \omega_{\mu\nu}
\left[ 4g^{\mu\nu}\left({\cal E}^{\scr{int}}-{\cal E}^{\scr{pot}}\right)
+ 2{\cal E}_{\scr{(der.)}}^{\mu\nu}\right], 
\end{eqnarray}
%
where the ${\cal E}_{\scr{(der.)}}^{\mu\nu}$ results from the
variation of $q$ factors in vertices of $\Phi^{\pi N}$.
Alternatively, performing variation of
$\Phi^{\pi N}$ accordingly to the canonical variation rules
(\ref{var-Phi-component}) and equating the result to expression
(\ref{xiLP-calc-pi}), we arrive at another identity
%
\begin{eqnarray}\label{LP-dPhi-pi}
\hspace*{-8mm}
&&\Tr\int \dpi{p}
\left[\left(
\Ldt_N\frac{\partial\Re\Gr^R}{\partial p_\mu} +
\Re\Se^R\frac{\partial\F_N}{\partial p_\mu}
\right) +  
\frac{1}{2}
\left(
\Ldt_\pi\frac{\partial\Re\Delta^R}{\partial p_\mu} +
\Re\Pi^R\frac{\partial\F_\pi}{\partial p_\mu}
\right)\right]  
\nonumber\\
&=& 
2g^{\mu\nu}\left({\cal E}^{\scr{int}}-{\cal E}^{\scr{pot}}\right)
+ {\cal E}_{\scr{(der.)}}^{\mu\nu}, 
\end{eqnarray}
%

The r.h.s. of the consistency relation for
energy--momentum conservation (\ref{epsilon-invk}) now reads
%
\begin{eqnarray}
\label{K-pN}
K^{\nu} =
\mbox{Tr}
\int 
\frac{p^\nu \di^4 p}{(2\pi )^4}
&&\left[
\left(
\Pbr{\Re\Se^R,\F_N} - \Pbr{\Re\Gr^R,\Ldt_N} 
\right)
\right.
\cr 
+ &&\frac{1}{2}
\left.
\left(
\Pbr{\Re\Pi^R,\F_\pi} - \Pbr{\Re\Delta^R,\Ldt_\pi} 
\right)
\right], 
\end{eqnarray}
%
By means of identity (\ref{f[]dp}) it is transformed to the form 
%
\begin{eqnarray}
\label{K1-pN}
K^{\nu} &=&
- \partial^\mu \mbox{Tr}
\int 
\frac{\di^4 p}{(2\pi )^4}
p^\nu \left[
\left(
\Re\Se^R  \frac{\partial\F_N}{\partial p^\mu} 
+ 
\Ldt_N \frac{\partial\Re\Gr^R}{\partial p^\mu}  
\right)\right.
\cr 
&& \hspace*{30mm} + \frac{1}{2}
\left.\left(
\Re\Pi^R  \frac{\partial\F_\pi}{\partial p^\mu} 
+ 
\Ldt_\pi \frac{\partial\Re\Delta^R}{\partial p^\mu}  
\right)
\right] 
\cr 
&&- 
\mbox{Tr}
\int \frac{\di^4 p}{(2\pi )^4}
\left[
\left(\Re\Se^R \partial^\nu \F_N + \Ldt_N \partial^\nu \Re\Gr^R\right)
\right.
\cr 
&& \hspace*{20mm} + \frac{1}{2}
\left.
\left(\Re\Pi^R \partial^\nu \F_\pi + \Ldt_\pi \partial^\nu \Re\Delta^R\right)
\right] .   
\end{eqnarray}
%
Now, applying identities (\ref{X-dPhi-pi}) and
(\ref{LP-dPhi-pi}) to the r.h.s. of Eq. (\ref{K1-pN}), we obtain 
%
\begin{eqnarray}
\label{K2-pN}
 K^{\nu} &=&
\partial^{\nu}
\left(
{\cal E}^{\scr{pot}} - {\cal E}^{\scr{int}} 
\right)-\partial_\mu{\cal E}_{\scr{(der.)}}^{\mu\nu}
\end{eqnarray}
%
which is precisely needed for the local conservation of the
Noether energy--momentum.


\end{fmffile}

\end{document}